\documentclass[usenatbib]{mnras}
\usepackage{newtxtext,newtxmath}

\usepackage[T1]{fontenc}

\DeclareRobustCommand{\VAN}[3]{#2}
\let\VANthebibliography\thebibliography
\def\thebibliography{\DeclareRobustCommand{\VAN}[3]{##3}\VANthebibliography}

\usepackage{graphicx}	
\usepackage{amsmath}	
\usepackage{caption}
\usepackage{subcaption}
\usepackage{booktabs}
\usepackage{makecell}
\usepackage{array}
\usepackage{soul}

\title[Tidal Feature Detection with SSL]{Detecting Galaxy Tidal Features Using Self-Supervised Representation Learning}

\author[A. Desmons et al.]{
Alice Desmons,$^{1}$\thanks{E-mail: a.desmons@unsw.edu.au}
Sarah Brough,$^{1}$
Francois Lanusse$^{2}$
\\
$^{1}$School of Physics, University of New South Wales, NSW 2052, Australia\\
$^{2}$AIM, CEA, CNRS, Universit\'e Paris-Saclay, Universit\'e Paris Diderot, Sorbonne Paris Cit\'e, F-91191 Gif-sur-Yvette, France\\
}

\date{Accepted 2024 June 04. Received 2024 June 04; in original form 2023 August 15}

\pubyear{2022}

\begin{document}
\label{firstpage}
\pagerange{\pageref{firstpage}--\pageref{lastpage}}
\maketitle

\begin{abstract}
Low surface brightness substructures around galaxies, known as tidal features, are a valuable tool in the detection of past or ongoing galaxy mergers, and their properties can answer questions about the progenitor galaxies involved in the interactions. The assembly of current tidal feature samples is primarily achieved using visual classification, making it difficult to construct large samples and draw accurate and statistically robust conclusions about the galaxy evolution process. With upcoming large optical imaging surveys such as the Vera C. Rubin Observatory’s Legacy Survey of Space and Time (LSST), predicted to observe billions of galaxies, it is imperative that we refine our methods of detecting and classifying samples of merging galaxies. This paper presents promising results from a self-supervised machine learning model, trained on data from the Ultradeep layer of the Hyper Suprime-Cam Subaru Strategic Program optical imaging survey, designed to automate the detection of tidal features. We find that self-supervised models are capable of detecting tidal features, and that our model outperforms previous automated tidal feature detection methods, including a fully supervised model. An earlier method 
applied to real galaxy images achieved 76\% completeness for 22\% contamination, while our model achieves considerably higher (96\%) completeness for the same level of contamination. We emphasise a number of advantages of self-supervised models over fully supervised models including maintaining excellent performance when using only 50 labelled examples for training, and the ability to perform similarity searches using a single example of a galaxy with tidal features.

\end{abstract}




\section{Introduction}
\label{sec:intro}
\begin{figure}
    \vspace{2em}%
    \centering
    \begin{subfigure}[t]{0.45\columnwidth}
        \centering
        \includegraphics[width=\columnwidth]{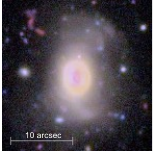}
    \end{subfigure}
    \hspace{0.1em}%
    \begin{subfigure}[t]{0.45\columnwidth}
        \centering
        \includegraphics[width=\columnwidth]{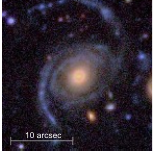}
    \end{subfigure}
    \vspace{0.5em}%
    \begin{subfigure}[b]{0.45\columnwidth}
        \centering
        \includegraphics[width=\columnwidth]{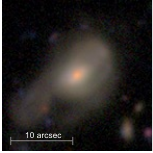}
    \end{subfigure}
    \hspace{0.1em}%
    \begin{subfigure}[b]{0.45\columnwidth}
        \centering
        \includegraphics[width=\columnwidth]{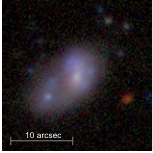}
    \end{subfigure}
    \caption{Example of galaxies with tidal features from HSC-SSP (\textit{gri}-band cutout images). Top row from left to right: shells, stream. Bottom row from left to right: asymmetric halo, and double nucleus.}
    \label{fig:examples}
\end{figure}

The currently accepted model of the Universe, known as the Lambda Cold Dark Matter ($\Lambda$CDM) Cosmological Model, postulates that galaxies evolve through a process which is referred to as the `hierarchical merger model’, wherein the growth of the universe's highest-mass galaxies is dominated by merging with lower-mass galaxies (e.g. \citealt{Lacey1994NBodyMergeRate, Cole2000Hierarchical, Robotham2014GAMAClosePair, Martin2018MergeMorphTransform}). During the merging process, the extreme gravitational forces involved cause stellar material to be pulled out from the galaxies, forming diffuse non-uniform regions of stars in the outskirts of the galaxies, known as tidal features (e.g. \citealt{Toomre1972BridgeTails}). Examples of these features from the optical Hyper Suprime-Cam Subaru Strategic Program (HSC-SSP; \citealt{Aihara2018HSCSurveyDesign}) are shown in Figure \ref{fig:examples}. These tidal features contain information about the merging history of the galaxy, and can thus be used to study the galaxy evolution process. Using tidal features to detect merging galaxies has a number of advantages over other methods such as spectroscopically detected galaxy close pairs. Not only do they remain observable significantly longer than close pairs (a few Gyr compared to $\sim$~600~Myr; \citealt{Lotz2011MajorMinorRate,Hendel2015TidalDebOrbit,Huang2022HSCTidalFeatETG}) but they can also be used to identify  mergers where a companion galaxy has already been ripped apart or is too low mass to be detected spectroscopically. Hence, using tidal features to study galaxy evolution can provide important observational confirmation on the contribution of low-mass galaxies to the merging process (e.g. \citealt{Johnston2008GalFormStellarHalo}).

In order to draw accurate and statistically robust conclusions about this evolution process, we require a large sample of galaxies exhibiting tidal features. This is difficult to achieve due to the extremely low surface brightness of tidal features, which can easily reach $\mu_{r}\geq$~27~mag~$\rm{arcsec}^{-2}$. The limiting surface brightness of wide-field optical astronomical surveys often do not reach these depths and as a consequence, many tidal features will not be identified simply because the features are not visible. This not only causes tidal feature incidence measures to be too low, but also increases the work required and number of images that need to be classified to assemble a sample with a significant number of galaxies with tidal features. With the next generation of wide-field optical imaging surveys reaching new limiting depths, such as the Vera C Rubin Observatory's Legacy Survey of Space and Time (LSST; \citealt{Ivezic2019LSST}) which is predicted to reach $\mu_{r}\sim$~30.3~mag~$\rm{arcsec}^{-2}$ \citep{Martin2022TidalFeatMockIm}, assembling a statistically significant sample of galaxies with tidal features is becoming more feasible. One challenge associated with surveys like LSST, due to commence in 2025 and run for 10 years, is the amount of data predicted to be released, with LSST predicted to output over 500 petabytes of imaging data including billions of galaxies \citep{Ivezic2019LSST}. Current tidal feature detection and classification is primarily achieved through visual identification (e.g. \citealt{Tal2009EllipGalTidalFeat, Sheen2012PostMergeSigs, Atkinson2013CFHTLSTidal, Hood2018RESOLVETidalFeat, Bilek2020MATLASTidalFeat, Martin2022TidalFeatMockIm}), but billions of galaxies are virtually impossible to classify visually by humans, even using large community based projects such as Galaxy Zoo \citep{Lintott2008GalZoo, Darg2010GalZooFracMarge}, and hence we are in urgent need of tools that can automate this classification task and isolate galaxies with tidal features.

Recent years have seen a steady increase in the use of machine learning for these types of data-intensive astrophysical classification tasks \citep{Huertas-Company2023DLAstroReview}. Such tasks have included grouping galaxies according to colour and morphology using both supervised Convolutional Neural Networks (CNNs; e.g. \citealt{Hocking2018UnsupGalMorph,Martin2020UnsupMorphClass}) and self-supervised networks (e.g. \citealt{Hayat2021SSMLAstroIms}), identifying the formation pathways of galaxies using a supervised CNN (e.g. \citealt{Cavanagh2020DeepLearnBars,Diaz2019CNNGalFormProcess}), and identifying new strong gravitational lens candidates using a self-supervised network (e.g. \citealt{SteinSep2021SelfSupSimfull,Stein2022SelfSupGravLens}). A number of works have also focused on classifying mergers and non-mergers using both random forest algorithms (e.g. \citealt{Snyder2019IllustrisAutoMergerClass}) and supervised CNNs (e.g. \citealt{Pearson2019DeepLearnMergers,SuelvesPearson2023CNNSDSSMergers}). CNNs have even shown great potential for the identification and classification of low surface brightness features such as tidal features (e.g. \citealt{Walmsley2019CNNTidalFeat, Bickley2021CNNTidalIllustris, Dominguez2023MockTidalCNN, Gordon2024DecalsTFCNN}).

With the promising recent results of machine learning in galaxy classification tasks, we turn to machine learning to construct a model which can take galaxy images as input, convert them into representations - low-dimensional maps which preserve the important information in the image - and output a classification based on whether the galaxy exhibits tidal features. We are looking for a tool which can perform classification and be told what features to look for, so an unsupervised machine learning model is not ideal. We also want a tool which does not require large labelled datasets for training, due to there being few such datasets of galaxies with tidal features and these being time-demanding to construct in themselves, so a supervised model is not ideal either. Instead we use a recently developed machine learning method that is essentially a middle-point between supervised and unsupervised learning, known as self-supervised machine learning (SSL; \citealt{He2019UnsupMomentContrast,ChenT2020SelfSup,ChenX2020MomentContrastive,ChenT2020ContrastiveFrame,ChenX2020SimSiam}). Such models do not require labelled data for the training of the encoder, which learns to transform images into meaningful low-dimensional representations, but can perform classification when paired with a linear classifier and a small labelled dataset. Instead of labels, SSL models rely on augmentations (e.g. image rotation, noise addition, PSF blurring) being applied to the encoder training dataset, to learn under which conditions the output low-dimensional representations should be invariant. These types of models have been successfully used for a variety of astronomical applications \citep{Huertas-Company2023ContLearn} including classification of galaxy morphology (e.g. \citealt{Walmsley2022SSMLGalMorph,Wei2022SSMLMorph,Ciprijanovic2023SSMLCrossSurv,Vega-Ferrero2023SSMLJWST}), clustering of galaxies according to age, metallicity, and velocity (e.g. \citealt{Sarmiento2021MaNGASSL}), estimation of black hole properties (e.g. \citealt{Shen2022SSMLBlackHole}), radio galaxy classification (e.g. \citealt{Slijepcevic2022SSMLRadio, Slijepcevic2023SSMLRadio}), and classification of solar magnetic field measurements (e.g. \citealt{Lamdouar2022SSMLSolar}). 
Another benefit to self-supervised models, other than training on unlabelled data, is the computational cost, self-supervised models have been shown to need only 1\% of the computational power needed by supervised models \citep{Hayat2021SSMLAstroIms}. Self-supervised models are also much easier to adapt to perform new tasks, and apply to datasets from new data releases or different astronomical surveys \citep{Ciprijanovic2023SSMLCrossSurv}, making this kind of model perfect for our goal of applying a tool developed using HSC-SSP data to future LSST data and potentially other future large imaging surveys such as Euclid \citep{Borlaff2022Euclid} and the Nancy Grace Roman Space Telescope \footnote{\url{https://roman.gsfc.nasa.gov/}} \citep{Spergel2015Roman}. \citet{Hayat2021SSMLAstroIms} and \citet{Stein2022SelfSupGravLens} show that once the encoder section of a self-supervised model has been trained one can conduct a simple similarity search to find objects of interest, or apply a linear classifier directly onto the encoder's outputs to separate the data into a variety of classes.

In this paper, we demonstrate that SSL can be used to detect tidal features amidst a dataset of thousands of otherwise uninteresting galaxies using $\sim$50,000 $grizy$-band Hyper Suprime-Cam Subaru Strategic Program (HSC-SSP; \citealt{Aihara2018HSCSurveyDesign}) 128$\times$128 pixel galaxy images. We also show the advantage of using a self-supervised model, as opposed to a supervised model, for the detection of merging galaxies, particularly in the regime of fewer labels, using the dataset of $\sim$ 6000 Sloan Digital Sky Survey (SDSS; \citealt{York2000SDSS}) galaxies constructed by \citet{Pearson2019DeepLearnMergers}. In Section \ref{sec:Methods} we detail data and sample selection, as well as the model architecture, including the augmentations we apply to the data. Section \ref{sec:results} details our results including our model's ability to detect tidal features in HSC-SSP data and a comparison of our model's performance to a supervised model when used for the detection of merging galaxies in SDSS images. In Section \ref{sec:disc} we compare our results with those in the literature and present our conclusions.

\section{Methods}
\label{sec:Methods}
\subsection{Data Sources and Sample Selection}
\label{sec:data}
For this work we use two separate datasets sourced from two different surveys. The first dataset is assembled from HSC-SSP galaxies and is used to show the potential of self-supervised machine learning models for the detection of tidal features. The second dataset consists of SDSS galaxies and is used to compare the performance of this self-supervised model with an earlier supervised model for the detection of merging galaxies.

\subsubsection{HSC-SSP Dataset}
\label{sec:HSC}
The HSC-SSP dataset used for this work is sourced from the Ultradeep (UD) layer of the HSC-SSP Public Data Release 2 (PDR2; \citealt{Aihara2019HSCSecondData}) for deep galaxy images. The HSC-SSP survey \citep{Aihara2018HSCSurveyDesign} is a three-layered, \textit{grizy}-band imaging survey carried out with the Hyper Suprime-Cam (HSC) on the 8.2m Subaru Telescope located in Hawaii. During the development of this project the HSC Public Data Release 3 (\citealt{Aihara2022HSCThirdData}) became available. Although this is an updated version of the HSC-PDR2, we do not use this data due to the differences in the data treatment pipelines. HSC-PDR2 has been widely tested for low surface brightness studies (e.g. \citealt{Huang2018HSCStellarHalo,Huang2020WeakLens,Li2022MassiveGalOutskirt,MartinezLombilla2023GAMAIntragroupLight}) and fulfils the requirements for our study. The HSC-SSP survey comprises of three layers: Wide, Deep, and Ultradeep which are observed to varying surface brightness depths. We use the Ultradeep field, which spans an area of $3.5$~deg$^{2}$ and reaches $\mu_{r}\sim$ 29.82~mag arcsec$^{-2}$ \citep{MartinezLombilla2023GAMAIntragroupLight}, a surface brightness depth faint enough to detect tidal features. We use HSC-SSP data not only for its depth, allowing us to detect tidal features, but also due to its similarity to LSST data. HSC-SSP data are reduced using the LSST pipeline \citep{Bosch2018HSCPipeline, Aihara2019HSCSecondData} and since the two surveys produce similar data, it will be more straightforward to adapt our SSL model and train it on LSST data once it is released. The HSC-SSP PDR2 has a median $\it{i}$-band seeing of 0.6 arcsec and a spatial resolution of 0.168 arcsec per pixel.

We assemble our initial unlabelled dataset of $\sim$50,000 galaxies by parsing objects in the HSC-SSP PDR2 Ultradeep database using an SQL search and only selecting objects which satisfy a pre-defined criteria. We describe our criteria, their definitions, and our reasoning for choosing them in Table \ref{tab:SQL_crit}. We filter our sample of 50,000 galaxies to remove repeat objects in the dataset by removing objects whose declinations and right ascensions are too close together (within a 5 arcsec radius circle), leaving us with a dataset of $\sim$~44,000 unlabelled objects. The galaxies in this dataset have a median \textit{i}-band magnitude \textit{i}$~=~$19.4 mag and median photometric redshift \textit{z}$~=~$0.29. We access the HSC-SSP galaxy images using the \texttt{Unagi} \textsc{Python} tool \citep{Huang2019Unagi} which, given a galaxy’s right ascension and declination, allows us to create multi-band ‘HSC cutout’ images of size 128~$\times$~128 pixels ( 21~$\times$~21 arcsecs), centred around each galaxy. Each cutout is downloaded in five ($g,~r,~i,~z,~y$) bands.

\begin{figure}
    \centering
	\includegraphics[width=0.99\columnwidth]{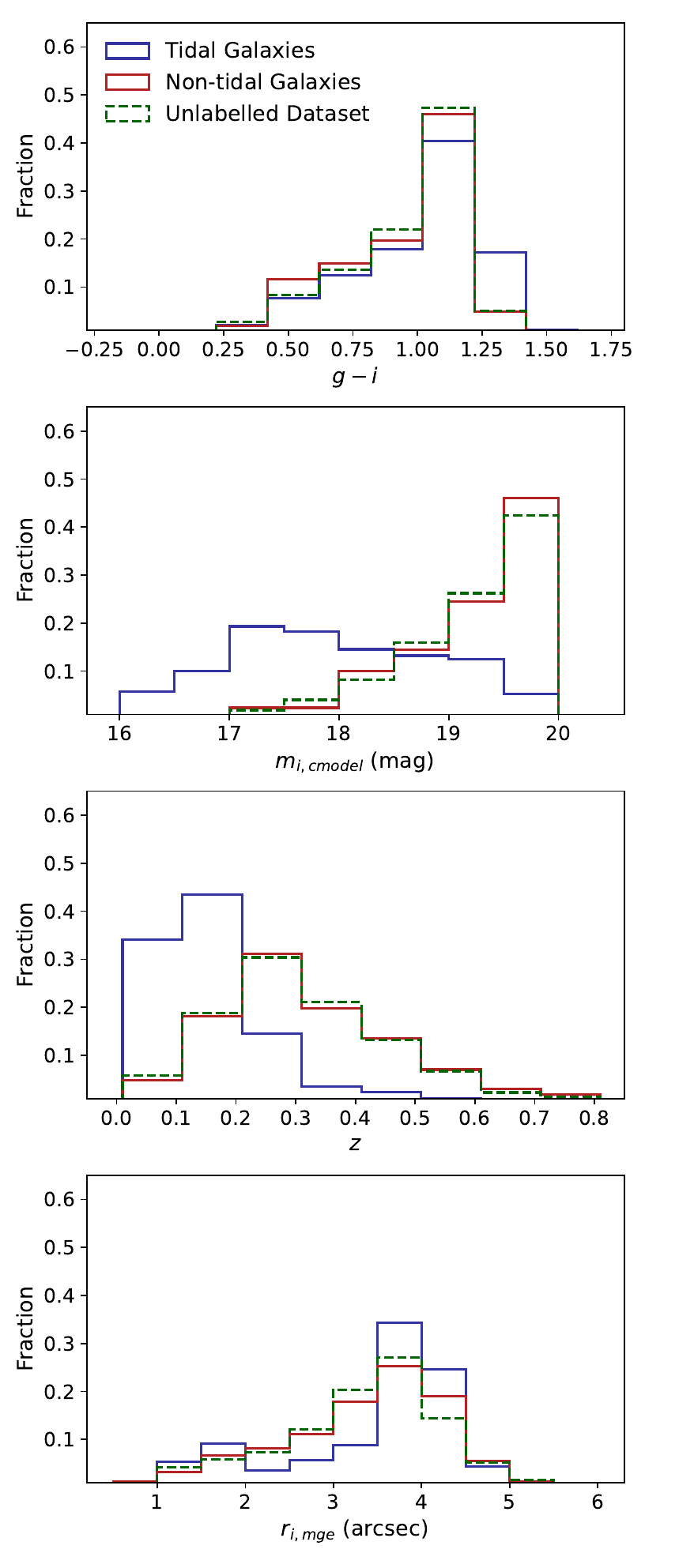}
    \caption{Distribution of galaxy $\it{g-i}$ colour (top), $\it{i-}$band magnitude (middle top), redshift (middle bottom), and radius (bottom) for the labelled datasets of galaxies with known tidal features (blue) and without tidal features (red) and the large unlabelled dataset (dashed green).}
    \label{fig:Data_prop}
\end{figure}

As mentioned in Section \ref{sec:intro} self-supervised networks encode images into meaningful lower-dimensional representations but cannot inherently perform classification tasks. To perform the classification, we will be using a linear classifier which takes in the encoded representations as input and classifies galaxies based on the presence of tidal features. To train this linear classifier we require a small labelled dataset of galaxies with and without tidal features. For the labelled dataset of galaxies with tidal features we use the HSC-SSP PDR2 Ultradeep dataset assembled by \citet{Desmons2023GAMA} from a $\sim$2.4 deg$^{2}$ crossover region of the Galaxy And Mass Assembly (GAMA; \citealt{Driver2011GAMADataRel}) survey with the HSC-SSP Ultradeep regions \citep{Desmons2023GAMA}. GAMA is an apparent-magnitude limited survey ($r~\leq~19.8$ mag) and has a median redshift $z~\sim~0.2$ (\citealt{Liske2015GAMADR2}). The galaxies were selected from a volume-limited sample with spectroscopic redshift limits 0.04~$\leq$~\textit{z}~$\leq$~0.2 and stellar mass limits 9.50~$\leq$~log$_{10}$($M_{\star}$/M$_{\odot}$)~$\leq$~11.00 and have \textit{i}-band magnitudes in the range 12.8~$\leq$~\textit{i}~$\leq$~21.6~mag (median \textit{i}$~=~$18.3 mag). This results in a labelled dataset of 211 galaxies with tidal features. To increase the size of our tidal feature training sample we classified additional galaxies from our HSC-SSP PDR2 unlabelled dataset of $\sim$~44,000 objects, according to the classification scheme outlined in \citet{Desmons2023GAMA}. The 380 galaxies in the final tidal feature training sample have \textit{i}-band magnitudes in the range 15.8~$\leq$~\textit{i}~$\leq$~20.1~mag (median \textit{i}$~=~$17.9 mag) and photometric redshifts in the range 0.02~$\leq$~\textit{z}~$\leq$~1.01 (median \textit{z}$~=~$0.14; as there are only 3 galaxies in the range 0.80~$\leq$~\textit{z}~$\leq$~1.01 it is uninformative to include them in Figure \ref{fig:Data_prop}). The classification was performed by the first author. The labelled dataset of galaxies without tidal features is assembled by randomly selecting galaxies from the unlabelled dataset and has median \textit{i}-band magnitude \textit{i}$~=~$19.4 mag and median photometric redshift \textit{z}$~=~$0.30.

We demonstrate the properties of our galaxy samples in Figure \ref{fig:Data_prop} which shows the distribution of colour, brightness, photometric redshift, and $i-$band radius in our unlabelled dataset ($\sim$44,000 galaxies) and our labelled datasets with tidal features ($\sim$400 galaxies) and without tidal features ($\sim$400 galaxies).

We found the Kron radii listed in the available HSC-SSP PDR2 catalogues to be unreliable and affected by bad segmentation and "bright galaxy shredding" where bright ($i~<~19$ mag) galaxies are deblended into multiple objects \citep{Aihara2018HSCSurveyDesign}. Galaxy shredding affects as much as 15\% of bright HSC-SSP galaxies, particularly late-type galaxies, causing Kron radius measurements to be significantly underestimated. Instead we calculate $i-$band galaxy radii by using the \texttt{find\_galaxy} function from the MGE fitting method and software developed by \citet{Cappellari2002MGE}. This method uses the weighted first and second moments of the intensity distribution to fit an ellipse to the galaxy. We then obtain the radius using $r_{i,MGE}=\sqrt{a~\times~b}$ where a and b are the lengths of the semi-major and semi-minor axes of the ellipse.

The galaxies from the GAMA dataset tend to be brighter and have lower redshifts as well as slightly larger apparent radii than the galaxies from the unlabelled and labelled non-tidal feature datasets, which are drawn from the broader HSC-SSP Ultradeep regions. As a deeper imaging survey, HSC-SSP reaches fainter absolute magnitudes, leading to a broader redshift, brightness, and resulting apparent radius range.

We use an equal number of galaxies with and without tidal features for model training and testing to prevent the model from learning an accidental bias against the category with fewer images. Our final labelled sample contains 760 galaxies, 380 with tidal features, labelled 1, and 380 without, labelled 0. Usually a labelled dataset is split into 80\%, 10\%, and 10\% for training, validation, and testing respectively. However, our labelled dataset is small and we want to maximise the number of galaxies in our testing dataset such that the model can be evaluated accurately. Hence we split our labelled dataset set into training, validation, and testing datasets composed of 600 (79\%), 60 (8\%), and 100 (13\%) galaxies respectively. 

\begin{table*}
    \small
    \centering
    \caption{Criteria used in our SQL search to select our training sample. An `x' indicates the cut was applied to each band $(g,r,i,z,y)$.}
    \label{tab:SQL}
    \renewcommand{\arraystretch}{1.4}
    \begin{tabular}{>{\ttfamily}p{6cm} p{9.5cm} }
        \toprule
        \thead{Criteria} & \thead{Definition and Reasoning} \\
        \midrule
        15 < i\_cmodel\_mag < 20 & $i-$band flux (mag) from the final cmodel fit. We set a faint magnitude limit of 20 mag to ensure that objects are bright enough for tidal features to be visible. \\
        x\_inputcount\_value $\geq$ 3 & The number of exposures available in a given band. We only select images which have at least 3 exposures in each band $(g,r,i,z,y)$ to ensure galaxies have full depth and colour information. \\
        NOT x\_pixelflags\_bright\_objectcenter \newline NOT x\_pixelflags\_bright\_object & Flags objects which are affected by bright sources. We set this criterion to avoid selecting objects affected by bright stars. \\
        NOT x\_pixelflags\_edge & Flags objects which intersect the edge of the exposure region. Ensures that objects are fully visible and not cut off by the edge of a region. \\
        NOT x\_pixelflags\_saturatedcenter & Flags objects which have saturated central pixels. We set this to avoid selecting objects with saturated pixels. \\
        NOT x\_cmodel\_flag & Flags objects which have general cmodel fit failures. Ensures we only select objects with good photometry. \\
        i\_extendendness\_value > 0.5 & Provides information about the extendedness of an object, values $<$ 0.5 indicate stars. We set this $>$ 0.5 to ensure we only select galaxies and not stars. \\
        \bottomrule
    \end{tabular}
    \label{tab:SQL_crit}
\end{table*}

\subsubsection{SDSS Dataset}
\label{sec:SDSS}
Our second dataset, consisting of SDSS data release 7 \citep{Abazajian2009SDSSDR7} galaxies, was assembled by \citet{Pearson2019DeepLearnMergers} for the purpose of training a supervised network to classify merging and non-merging galaxies. This dataset contains $\sim$~10,000 non-merging galaxies and $\sim$~3000 merging galaxies which were selected from the \citet{Darg2010GalZooFracMarge,Darg2010GalZooPropMerge} catalogue, assembled from Galaxy Zoo \citep{Lintott2008GalZoo} classifications. All galaxies in this dataset have spectroscopic redshift limits 0.005~$\leq$~\textit{z}~$\leq$~0.1 and are in the stellar mass range 9.5~$\leq$~log$_{10}$($M_{\star}$/M$_{\odot}$)~$\leq$~12.0. SDSS imaging is incomplete at surface brightnesses fainter than $\mu_{r}\leq$~23~mag~arcsec$^{-2}$ \citep{Strauss2002SDSSSample} Due to this surface brightness limit being significantly brighter than the HSC-SSP dataset described above, only the brightest tidal features would be visible in these images i.e. tidal features fainter than $\mu_{r}=$~23~mag~arcsec$^{-2}$ are unlikely to be detected. This means that the merging galaxies in this sample were selected not based on the presence of tidal features, but rather based on whether galaxies showed obvious signs of mergers with at least two clearly interacting galaxies or significantly morphologically disturbed systems. To prevent the model learning an accidental bias against the category with fewer images we reduce the size of our non-merging dataset by randomly selecting only 3000 non-merging galaxies from the set of 10,000.
The SDSS dataset used to train and test our model now consists of 3000 merging galaxies, labelled 1, and 3000 non-merging galaxies, labelled 0. This gives us a sample of 6000 SDSS objects of which we obtain 256~$\times$~256 pixel images, downloaded in three ($g,~r,~i$) bands. For more detail about the construction and classification of the SDSS dataset we refer the reader to \citet{Pearson2019DeepLearnMergers}. When a labelled dataset is required for training (i.e. for the supervised model or the linear classifier) we split this dataset into 4800 (80\%), 600 (10\%), and 600 (10\%) galaxies to use for training, validation, and testing respectively.
{For the portion of the training of our self-supervised model which requires an unlabelled dataset, we increase the size of our SDSS dataset from 6000 to 50,000 galaxies by using augmentations, namely, rotating each image by random increments of 90 degrees.

The analysis presented in Section \ref{sec:results} is focused on comparing the performance of supervised and self-supervised models when a varying number of labels are used for training. When using fewer labels to train a model, we do not reduce the number of images shown to the model during training, but rather the number of unique images. For example, if the model trained on 80\% of the full SDSS training set is shown 4800 unique images, the model trained on 2\% of the labelled data is still shown 4800 images but only 120 of these images are unique. This ensures that the change in model performance is indeed due to the number of unique labelled examples, and not due to the model being shown fewer images.

\subsection{Image Pre-processing and Augmentations}
\label{sec:augs}
Before the images are augmented and fed through the model we apply a pre-processing function to normalise the images. For the HSC-SSP images this is done by taking a subsample of 1000 galaxies from our unlabelled training sample and calculating the standard deviation $\sigma_{\text{pixel count}}$ for each band $(g,r,i,z,y)$ of this subsample using the median absolute deviation. The entire sample is then normalised by dividing each image band by the corresponding $3\sigma$ and then taking the hyperbolic sine of this. In this step we also define an `unnormalising' function which does the inverse of the normalising function, allowing us to retrieve the original unormalised images if needed. The SDSS images we obtained from \citet{Pearson2019DeepLearnMergers} were in PNG format with pixel values between 0 and 255, having already been normalised using a modified version of the \citet{Lupton2004RGB} hyperbolic sine normalisation. To normalise these we simply divide each image band by 255.

Self-supervised networks work by encoding images into lower-dimensional representations and using augmented versions of the training images to learn under which transformations the encoded representations should be invariant. More specifically, these networks use contrastive loss \citep{Hadsell2006ContLoss} which is minimised for different augmentations of the same image, and maximised when the images are different. These augmentations are defined and applied before the images get processed by the network and are chosen based on the task at hand. In this project we are constructing a network to classify whether a galaxy possesses tidal features. This classification should be independent of the level of noise, or the orientation or position of the galaxy in the image. To achieve this we use the following augmentations:
\begin{itemize}
    \setlength\itemsep{0.8em}
    \item \textbf{Orientation:} We randomly flip the image across each axis (x and y) with 50\% probability.
    \item \textbf{Gaussian Noise}: We sample a scalar from $\mathcal{U}$(1,3) and multiply it with the median absolute deviation of each channel (calculated over 1000 training examples) to get a per-channel noise $\sigma_{c}$. We then introduce Gaussian noise sampled from $\sigma_{c}~\times~\mathcal{N}$(0,1) for each channel.
    \item \textbf{Jitter and Crop:} For HSC-SSP images we crop the 128~$\times$~128 pixel image to the central 109~$\times$~109 pixels before randomly cropping the image to 96~$\times$~96 pixels. Random cropping means the image centre is translated, or `jittered', along each respective axis by $i$, $j$ pixels where $i$, $j~\sim~\mathcal{U}$(-13,13) before cropping to the central 96~$\times$~96 pixels. For SDSS images we crop the 256~$\times$~256 pixel image to the central 72~$\times$~72 pixels before randomly cropping the image to 64~$\times$~64 pixels. We use 64~$\times$~64 pixel images when training with SDSS data as these are the image dimensions used by \citet{Pearson2019DeepLearnMergers} for their model. We use a smaller maximum centre translation ($i$, $j~\sim~\mathcal{U}$(-8,8)) for SDSS images due to their smaller size.
\end{itemize}

\subsection{Model Architecture}
\label{sec:mod_arch}
The model we utilise to perform classification of tidal feature candidates consists of two components; a self-supervised model used for pre-training, and a linear classifier used for classification. The self-supervised model receives galaxy images as input and encodes these into meaningful lower-dimensional representations. The linear classifier is a simple supervised model which takes in galaxy images, encodes them into representations using the trained self-supervised model, and outputs a binary classification. All models described below are built using the TensorFlow framework \citep{Abadi2016TensorFlow}.

\subsubsection{The Self-Supervised Architecture}
\label{sec:self_sup_arch}
For our task of classifying tidal feature candidates we use a type of self-supervised learning known as Nearest Neighbour Contrastive Learning of visual Representations (NNCLR; \citealt{Dwibedi2021NNCLR}). We closely follow \citet{Dwibedi2021NNCLR} in designing the training process for our model. A general schematic of the NNCLR  framework is shown in Figure \ref{fig:Model_arch} and we refer the reader to \citet{Dwibedi2021NNCLR} for a more detailed explanation of the approach.

Self-supervised models rely on augmentations to create different views of the same images. Given a sample of images \textbf{x}, a pair of images ($\mathrm{\mathbf{x_{i}}}$, $\mathrm{\mathbf{x_{j}}}$) is defined as positive when $\mathrm{\mathbf{x_{j}}}$ is an augmented version of image $\mathrm{\mathbf{x_{i}}}$, a positive pair can be expressed as ($\mathrm{\mathbf{x_{i}}}$, $\mathrm{\mathbf{x_i^+}}$). If  $\mathrm{\mathbf{x_{j}}}$ is not an augmented version of image $\mathrm{\mathbf{x_{i}}}$ the pair is negative and is expressed as ($\mathrm{\mathbf{x_{i}}}$, $\mathrm{\mathbf{x^{-}}}$). A network consisting of an encoder and projection head takes the images in \textbf{x} as input and outputs 128-dimensional representations of the images which we denote \textbf{z}. This network is trained to make the representations similar for positive pairs, and dissimilar for negative pairs, by using a contrastive loss function:
\begin{equation}
    L_{i} = -\log\left(\frac{\mathrm{exp(sim(\mathbf{z_i},\mathbf{z_i^+}))}}{\mathrm{exp(sim(\mathbf{z_i},\mathbf{z_i^+}))} + \sum\mathrm{\mathbf{_{z^-}}exp(sim(\mathbf{z_i},\mathbf{z^-}))}}\right)
    \label{eq:cont_loss}
\end{equation}
where sim(\textbf{a}, \textbf{b}) = \textbf{a}~$\cdot$~\textbf{b}~/~ ($\tau||\rm{\textbf{a}}||~||\rm{\textbf{b}}||$) is the cosine similarity between vectors \textbf{a} and \textbf{b}, normalised by the tunable "softmax temperature" $\tau$. This contrastive loss, or InfoNCE loss \citep{vandenOord2018RepLearnContrast}, is minimised when the similarity is high for positive pairs and low for negative pairs. Using this contrastive loss, self-supervised networks learn to make the representations similar for positive pairs, and dissimilar for negative pairs, and hence are able to cluster similar (or positive) samples together and push apart dissimilar (or negative) samples. These contrastive learning methods (e.g. SimCLR; \citealt{ChenT2020ContrastiveFrame}) rely only on differently augmented views of the same image to create positive pairs. As a consequence, objects which exhibit large variations but belong to the same class (e.g. galaxies with different types of tidal features) might not be linked using this type of method. NNCLR methods aim to resolve this issue by creating a more diverse set of positive pairs. Instead of defining positive pairs as ($\mathrm{\mathbf{x_{i}}}$, $\mathrm{\mathbf{x_i^+}}$) where $\mathrm{\mathbf{x_i^+}}$ is an augmented version of image $\mathrm{\mathbf{x_{i}}}$, NNCLR uses a queue of examples $\mathcal{Q}$, and defines $\mathrm{\mathbf{x_i^+}}$ as the nearest-neighbour of $\mathrm{\mathbf{x_{i}}}$ in the queue. The loss function for NNCLR models varies slightly from the InfoNCE loss and is defined as:
\begin{align*}
    L_{i}^{\rm{NNCLR}} =& -\log\left(\frac{\mathrm{exp(NN(\mathbf{z_i},\mathcal{Q})~\cdot~\mathbf{z_i^+}~/~\tau)}}{\sum\mathrm{\mathbf{_{k}}exp(NN(\mathbf{z_i},\mathcal{Q})~\cdot~\mathbf{z_k^+}~/~\tau)}}\right)\\
    &-\log\left(\frac{\mathrm{exp(NN(\mathbf{z_i},\mathcal{Q})~\cdot~\mathbf{z_i^+}~/~\tau)}}{\sum\mathrm{\mathbf{_{k}}exp(NN(\mathbf{z_k},\mathcal{Q})~\cdot~\mathbf{z_i^+}~/~\tau)}}\right)
    \label{eq:NNCLR_loss}
\end{align*}
where NN($\mathrm{\mathbf{x_{q}}}$, $\mathcal{Q}$) is the nearest neighbour operator defined as:
\begin{equation}
    \mathrm{NN(\mathbf{z},\mathcal{Q})} = \mathop{\arg \min}\limits_{i \in \mathcal{Q}}~||\mathrm{\mathbf{z}}-i||_2
\end{equation}

where $||\mathrm{\mathbf{x}}||_2$ represents $\mathrm{\mathbf{x}}$ $l_{2}$-normalised along the first axis, defined as:
\begin{equation}
    ||\mathrm{\mathbf{x}}||_2 = \sqrt{\sum_{k=1}^{n}|x_k|^2}
\end{equation}
The self-supervised model was trained using a temperature of 0.1 and a queue size of 10,000. We use ResNet-20 \citep{He2016deep} as our encoder. The architecture of the projection head is two fully connected layers of size 128, which use L2 kernel regularisation with a penalty of 0.0005. Each fully connected layer is followed by a batch-normalisation layer, and the first batch-normalisation layer is followed by ReLU activation. The model was compiled using the Adam optimiser \citep{Kingma2014AdamLoss} and trained for 25 epochs on our unlabelled dataset of $\sim$~44,000 HSC-SSP PDR2 galaxies or our dataset of 50,000 SDSS galaxies. Training was completed within $\sim$~30 minutes using a single NVIDIA A100 40GB GPU.

\begin{figure*}
  \includegraphics[width=0.9\textwidth,]{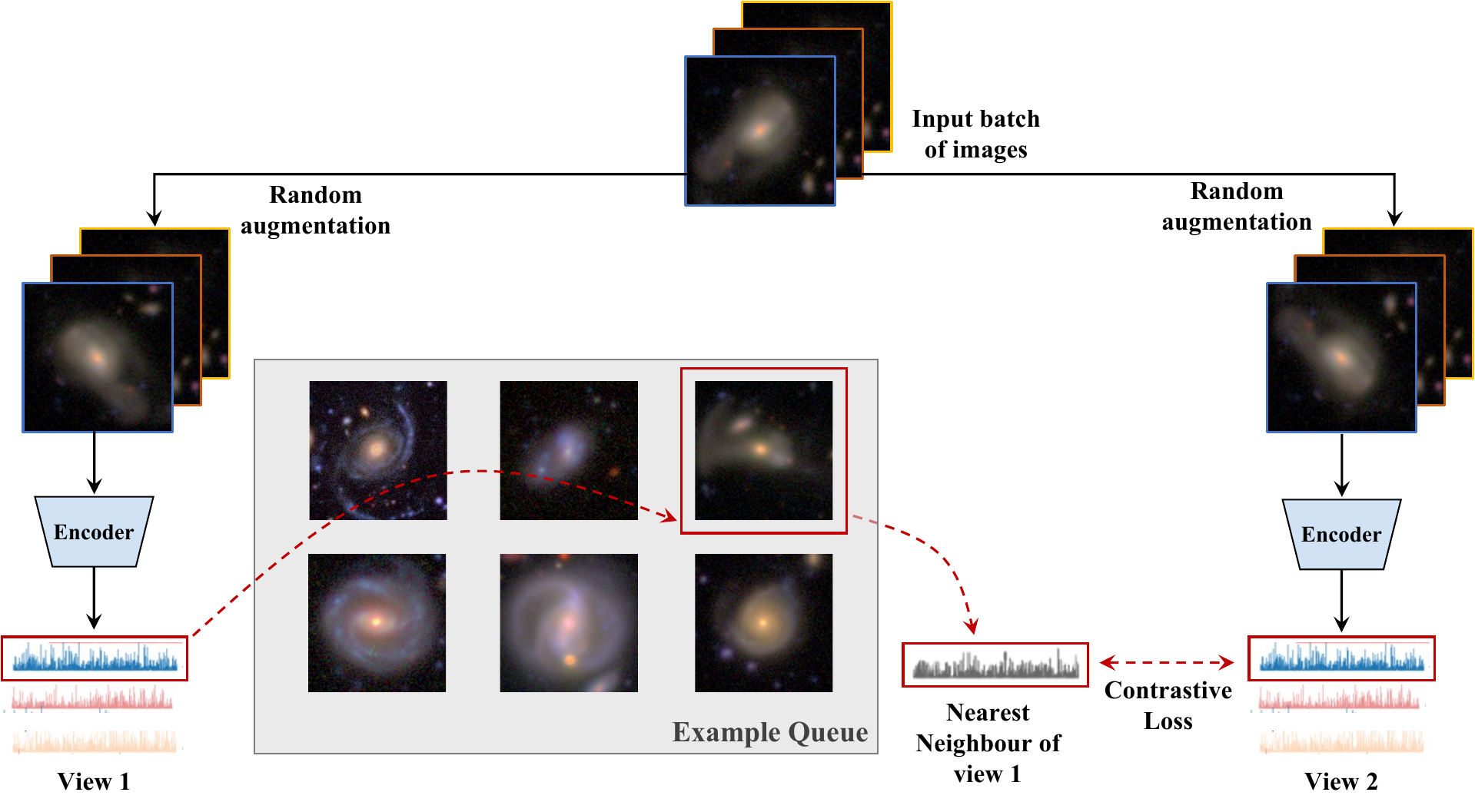}
  \caption{Illustration of the self-supervised model architecture. The model takes in a batch of images as input and creates two different views of the batch using random augmentations. Each view is encoded and the nearest neighbour of view 1 is located from the queue of example images. The loss between view 2 and the nearest neighbour of view 1 is calculated using a contrastive loss function which is minimised for similar pairs of images and maximised otherwise.}
  \label{fig:Model_arch}
\end{figure*}

\subsubsection{The Linear Classifier Architecture}
\label{sec:fine_tune_arch}
The second part of the model is a simple linear classifier which takes galaxy images as input and converts them to representations using the pre-trained self-supervised encoder. These encoded representations are passed through a fully connected layer with a sigmoid activation, which outputs a single number between 0 and 1. This fine-tuned model was compiled using the Adam optimiser \citep{Kingma2014AdamLoss} and a binary cross entropy loss. It was trained for 50 epochs using the labelled training set of 600 HSC-SSP galaxies or 4800 SDSS galaxies. Training was completed within $\sim$~1 minute using a single GPU.

\subsubsection{The Supervised Architecture}
\label{sec:sup_arch}
To draw conclusions about the suitability of self-supervised models for the detection and classification of tidal features, we compare our results with those of a fully supervised model. We do not construct this model from scratch, but instead use the model designed by \citet{Pearson2019DeepLearnMergers} to classify merging galaxies. This model consists of four convolutional layers, each followed by a Rectified Linear unit (ReLU) activation layer and a dropout layer with a dropout rate of 20\%. The first, second, and fourth convolutional layers also have max-pooling layers following the dropout layers. The model then has two dense layers, each followed by a ReLU activation layer and dropout layer. The final layer is a dense layer which has two neurons and softmax activation. We do not detail the dimensions of the network layers here but instead focus on the changes made to adapt the network to our data. For further detail on the architecture of this network we refer the reader to \citet{Pearson2019DeepLearnMergers}. The output layer was changed from two neurons with softmax activation, to a single neuron with sigmoid activation. The network was compiled using the Adam optimiser \citep{Kingma2014AdamLoss} with the default learning rate and loss of the network was determined using binary cross entropy.
When training the network on our HSC-SSP dataset we additionally changed the input image dimension from 64~$\times$~64 pixels with three colour channels, corresponding to the SDSS dataset configuration, to 96~$\times$~96 pixels with five colour channels. We do this because tidal features in deeper images can often be seen to extend around the galaxy and using 64~$\times$~64 pixel images can cause them to be cut-off. We train the supervised network from scratch using the labelled training set of 600 HSC-SSP galaxies or 4800 SDSS galaxies.

\subsection{Model Evaluation}
\label{sec:mod_eval}
There are a number of metrics used in the literature to evaluate the performance of machine learning models, depending on the task at hand. The purpose of our model is to partially automate the creation of large datasets of galaxies with tidal features by reducing the amount of data which has to be visually classified. If the top $N$ predictions will be visually classified, we want to maximise the number of true positives, or galaxies with tidal features, in the top $N$ predictions while minimising the number of false positives, or galaxies without tidal features. As such, in terms of model performance, we are primarily concerned with the true positive rate (also known as recall or completeness) and false positive rate (also known as fall-out or contamination). The true positive rate (TPR) ranges from 0 to 1 and is defined as:
\begin{equation}
    TPR = \frac{TP}{TP~+~FN}
\end{equation}
where $TP$ is the number of true positives (i.e. the number of galaxies with tidal features correctly classified by the model) and $FN$ is the number of false negatives (i.e. the number of galaxies with tidal features incorrectly classified by the model). The false positive rate (FPR) also ranges from 0 to 1 and is defined as:
\begin{equation}
    FPR = \frac{FP}{FP~+~TN}
\end{equation}
where $FP$ is the number of false positives (i.e. the number of galaxies without tidal features incorrectly classified by the model) and $TN$ is the number of true negatives (i.e. the number of galaxies without tidal features correctly classified by the model).

In addition to using the TPR for a given FPR to evaluate our model, we also use the area under the receiver operating characteristic (ROC) curve to evaluate performance. The ROC curve is a plot of the TPR against the FPR for a range of threshold values, which is the threshold between outputs being labelled positive or negative. For our model where galaxies with tidal features are labelled 1 and galaxies without are labelled 0, the threshold can take any value between 0 and 1 and determines whether outputs are classified as galaxies with or without tidal features. The area under the ROC curve (AUC) of a perfect model with TPR$~=~1$ and FPR$~=~0$ is unity, while a good model will have an AUC close to unity. A truly random model will have a ROC AUC of 0.5. 

\section{Results}
\label{sec:results}
In this section we first compare the performance of our self-supervised model with a supervised model, before exploring how our self-supervised model organises the galaxy images in representation space.

\subsection{Self-Supervised vs. Supervised Performance}
\label{sec:res_comp}

Figure \ref{fig:SDSS_AUC} illustrates the ROC AUC for a supervised and self-supervised network, evaluated on the test set, as a function of the number of labels used in training for our SDSS dataset. As shown, when the amount of training data is within the range of 600 labels to 4800 labels the models show very similar performances. However, in the regime of fewer labels, particularly when training with 300 or 120 unique labelled examples, both models show decreased performance but this decrease is significantly greater for the supervised model. When only 120 unique labelled training examples are used, the self-supervised model AUC only decreases to 0.85 while the supervised AUC drops to 0.77. This figure does not show that self-supervised models can be used in the detection of tidal features, because the criterion for galaxies being classed as merging in the SDSS dataset is mainly based on whether two clearly interacting galaxies could be seen in the images.

\begin{figure}
    \centering
	\includegraphics[width=0.9\columnwidth]{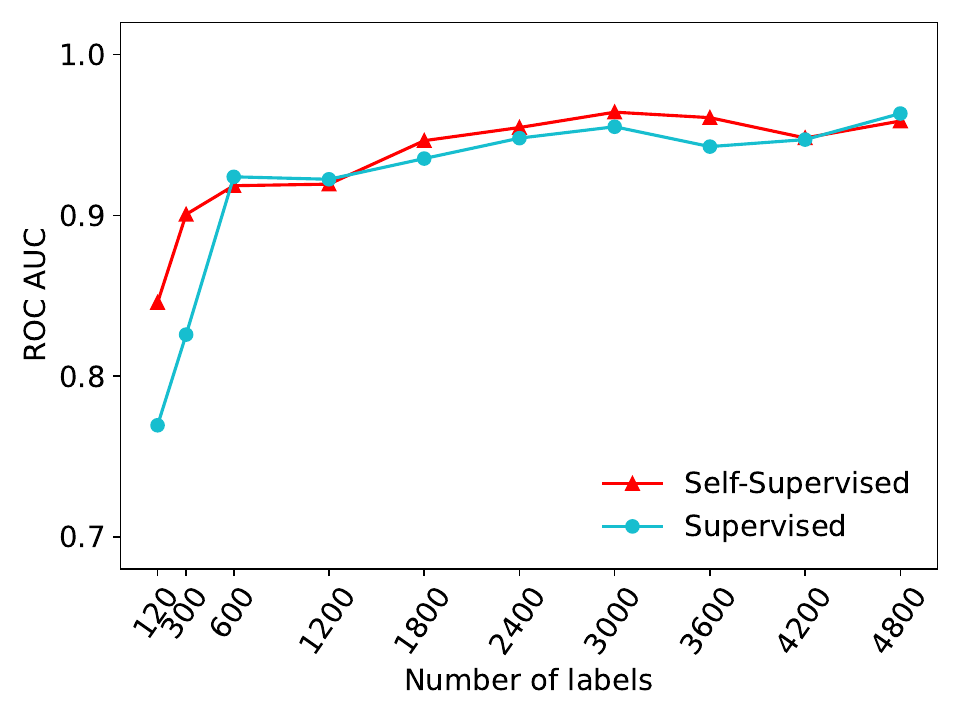}
    \caption{Test set ROC AUC as a function of the number of SDSS labels used for training for a supervised (blue) and self-supervised (red) model. Both models show similar performance in the high number of labels regime. For fewer training labels, the self-supervised model outperforms the supervised model.}
    \label{fig:SDSS_AUC}
\end{figure}

To show that self-supervised models can be used to detect galaxies with tidal features we rely on our HSC-SSP dataset, which reaches surface brightness depths sufficient to clearly identify tidal features. Figure \ref{fig:HSC_AUC} illustrates the testing set ROC AUC for a supervised and self-supervised network as a function of the number of labels used in training for our HSC-SSP dataset. Each point along the solid lines represents the ROC AUC averaged over ten runs using the same training, validation, and testing sets for each run. We average the ROC AUC over the 10 runs and remove outliers further than $3\sigma$ from the mean to remove any effects of network initialisation. It is important to note that in this figure, the number of labels for our rightmost data point, where 600 labels are used for training, is equivalent to the 10\% of labels data point for the SDSS results in Figure \ref{fig:SDSS_AUC}. Our SSL model maintains high performance across all amounts of labels used for training, having ROC~AUC~$=$~0.911~$\pm$~0.002 when training on the maximum number of labels and only dropping to ROC~AUC~$=$~0.89~$\pm$~0.01 when using only 50 labels for training. This is in contrast to the supervised model, which also maintains its performance regardless of label number, but only reaches ROC~AUC~$=$~0.867~$\pm$~0.004 when training on the maximum number and ROC~AUC~$=$~0.83~$\pm$~0.01 when using only 50 labels for training. This figure not only shows that a SSL model can be used for the detection of tidal features with good performance, but also that it performs consistently better than the supervised network regardless of the number of training labels. 

It seems unlikely that a purely supervised network would maintain such good performance with only 50 unique labelled training examples, however, upon inspection we found no problems with our dataset assembly or model training. Instead of only evaluating the model based on its performance with regards to the testing set, we also plot the results of the models based on the validation loss. We do this by choosing the `best' model from the ten runs for each number of labels, defined as the model which reaches the lowest validation loss at the end of training. Typically, a lower validation loss translates to a better model performance on the testing set. However, our validation and testing sets are very small, 60 and 100 galaxies respectively, making it hard to evaluate our models accurately using just one method. The points along the dashed lines in Figure \ref{fig:HSC_AUC} show the ROC AUC of the model which reached the lowest final validation loss for the given number of labels used for training. This shows that both models maintain similar ROC AUC when using more training labels, however, when using less than 300 labels for training, the supervised model begins to decrease in performance. When using as few as 50 labelled examples for training, the self-supervised model's ROC AUC remains stable around 0.9, whereas the supervised model's ROC AUC drops to 0.7. Although this may seem to contradict the results of the solid lines in Figure \ref{fig:HSC_AUC}, the two linestyles compare different methods of assessing the model. The solid lines are based only on the performance of the model on the testing set, whereas the dashed lines take into account the training process using the validation loss. The self-supervised network shows consistency in the ROC AUC regardless of which method is chosen to evaluate the model, while the supervised network appears to be more sensitive to the choice of method.

\begin{figure}
    \centering
	\includegraphics[width=0.95\columnwidth]{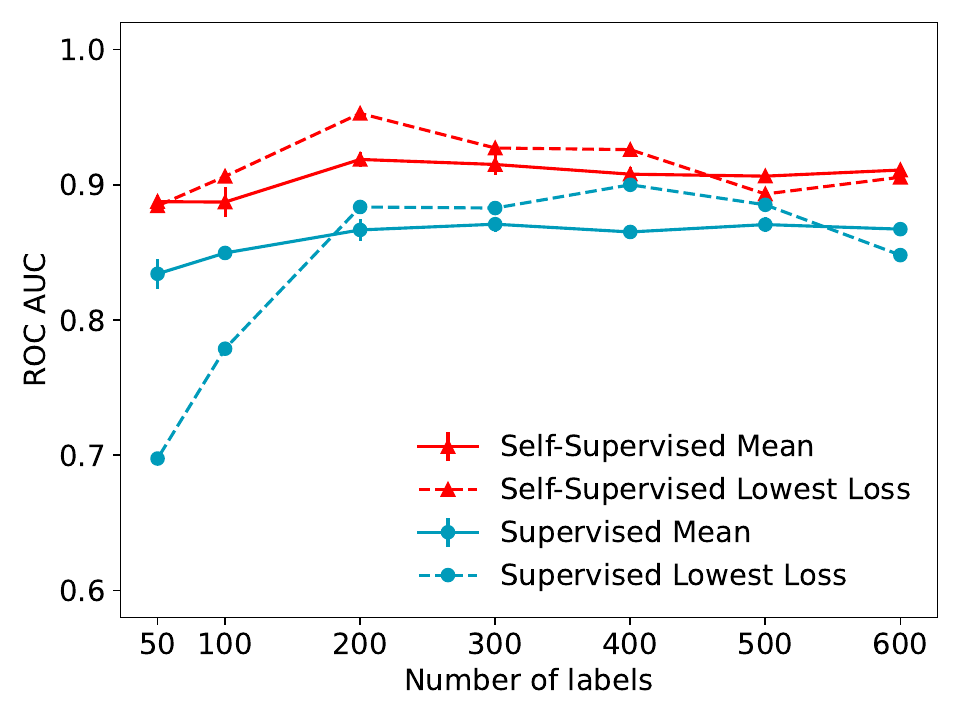}
    \caption{Test set ROC AUC as a function of the number of HSC-SSP labels used for training for a supervised (blue) and self-supervised (red) model. The points along the solid lines represent the average test set ROC AUC where each point is an average of ten runs. The points along the dashed lines show the ROC AUC for the model which reached the lowest final validation loss for the given number of training labels. When using the average ROC AUC the self-supervised model performs consistently better, but both models remain consistent for all number of labels. When using the model with the lowest validation loss both models show similar performance in the high number of labels regime. For fewer training labels, the self-supervised model outperforms the supervised model.}
    \label{fig:HSC_AUC}
\end{figure}

\begin{figure}
    \centering
	\includegraphics[width=0.95\columnwidth]{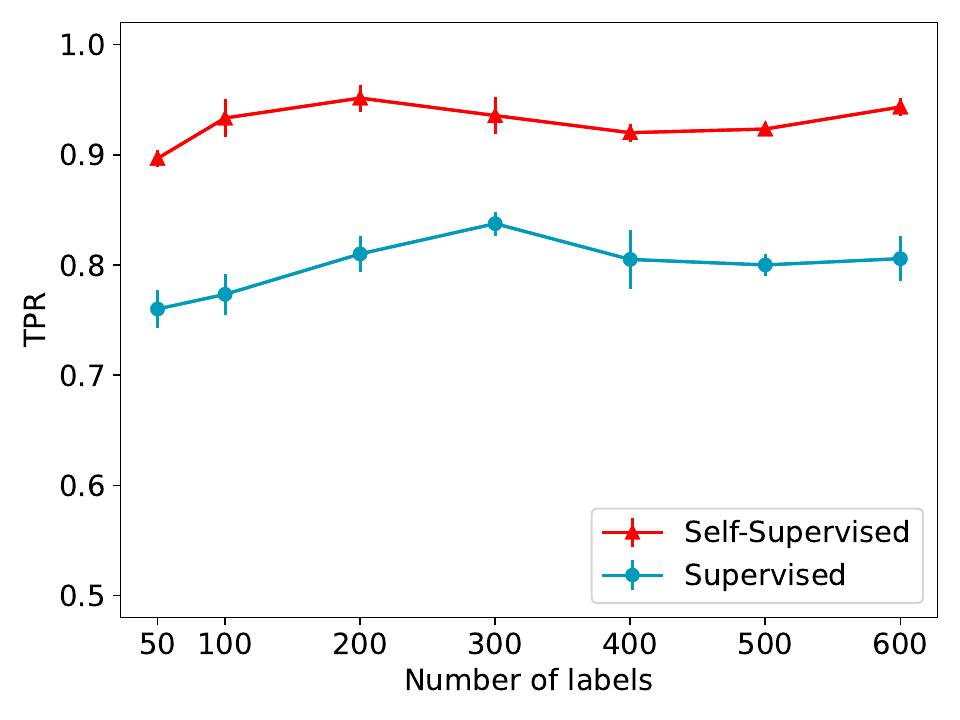}
    \caption{Average test set TPR when FPR = 0.2 as a function of the number HSC-SSP labels used for training for the supervised (blue) and self-supervised (red) model. Each point is an average of 10 runs. Both models show a slight decrease in TPR with decreasing training label number. The self-supervised model has a consistently higher TPR.}
    \label{fig:HSC_TPR}
\end{figure}

The solid lines in Figure \ref{fig:HSC_AUC} showed that our SSL model maintained its ROC AUC and performed consistently better than the supervised network regardless of the number of training labels. This result is also reflected in Figure \ref{fig:HSC_TPR} which shows the testing set TPR for FPR~$=$~0.2 for our supervised and self-supervised networks as a function of the number of labels used in training. Similar to the solid lines in Figure \ref{fig:HSC_AUC}, each point represents the TPR averaged over 10 runs, removing outliers further than $3\sigma$ from the mean. Figure \ref{fig:HSC_TPR} not only shows that the TPR for a given FPR is consistently higher for our self-supervised model than our supervised model but also that the self-supervised model mantains a high TPR regardless of the number of training labels, only dropping from TPR~$=$~0.94~$\pm$~0.01 at 600 training labels to TPR~$=$~0.90~$\pm$~0.01 with a mere 50 training labels. 

We do not specify an optimal threshold for the model in this paper as the threshold is application-dependent and the trade-off between the TPR and FPR should be chosen based on the science-case application of the model. Instead, the red curve in Figure \ref{fig:Best_ROC} shows the ROC curve for the model published to GitHub (see Section \ref{sec:disc}) to allow the user to choose their desired trade-off between TPR and FPR. This model is the linear classifier trained on 600 labels with the highest ROC AUC. The ROC curves for the other nine classifiers trained on 600 labels are shown in grey.

\begin{figure}
    \centering
	\includegraphics[width=0.95\columnwidth]{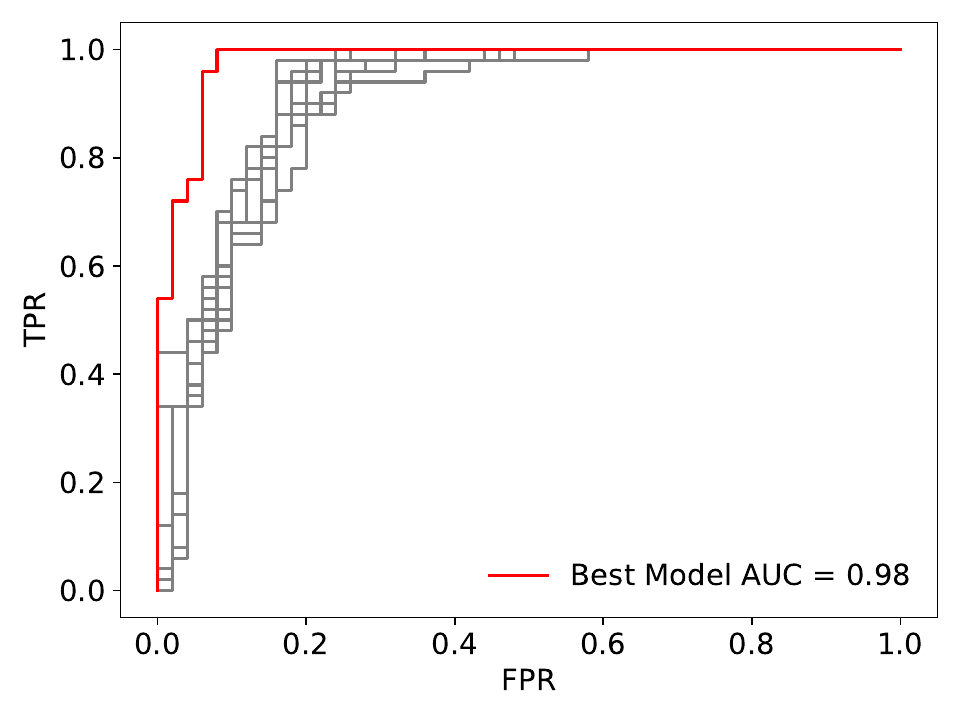}
    \caption{ROC curves for the ten self-supervised models trained on 600 labels. The model with the highest ROC AUC is shown in red, the other nine models are shown in grey. The model shown in red was an outlier and hence was not included in the average ROC AUC calculations in Figure \ref{fig:HSC_AUC}.}
    \label{fig:Best_ROC}
\end{figure}

\subsection{Detection of Tidal Features}
\label{sec:res_HSC}
\subsubsection{Similarity}
\label{sec:Similarity}
One advantage of self-supervised models over supervised models is the ability to use just one labelled example to find examples of similar galaxies from the full dataset. By using just one image from our labelled tidal feature dataset as a query image, and the encoded 128-dimensional representations from the self-supervised encoder, we can perform a similarity search that assigns high similarity scores to images which have similar representations to the query image. This is done using a similarity function which takes in the encoded representations of two images, $l_{2}$-normalises them along the first axis, and returns the reduced sum of the product of these normalised representations. This is demonstrated in Figure \ref{fig:HSC_sim_search} where we select two galaxies with tidal features from our training sample and perform a similarity search with the 44,000 unlabelled HSC-SSP galaxies. In Figure \ref{fig:HSC_sim_search} the query image is shown on the right alongside the 24 galaxies which received the highest similarity scores. This figure shows the power of self-supervised learning, where using only a single labelled example, we can find a multitude of other tidal feature candidates. The similarity search finds similar galaxies using only the 128-dimensional representations output by the encoder, which has not been explicitly trained to isolate galaxies with tidal features. While the similarity search function is useful to understand what type of information is stored in the encoded representations, it is not perfect. This is why we refine the process by training a classifier to specifically isolate galaxies with tidal features. Alongside the similarity scores in Figure \ref{fig:HSC_sim_search} we also display the classifier scores assigned to each galaxy by the classifier. We note that even galaxies without tidal features in this figure are assigned high classifier scores. However, this cannot be directly translated to the purity of a sample selected based on classifier output. The classifier scores are discussed in greater detail later in this section.

\begin{figure*}
  \includegraphics[width=0.99\textwidth,]{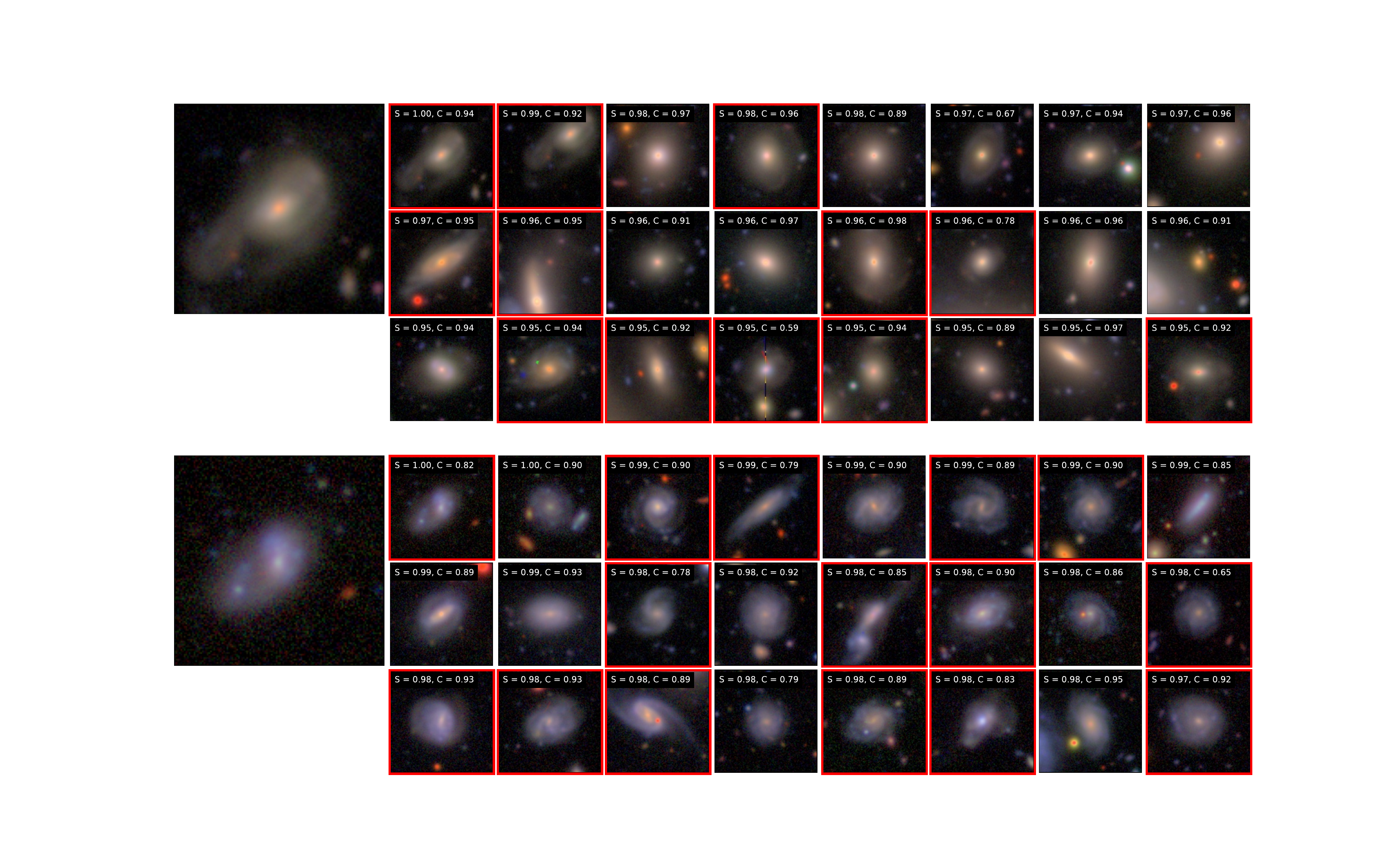}
  \caption{Results from a similarity search using two random galaxies with tidal features as query images. The two query galaxies are displayed on the left, alongside the top 24 galaxies with the highest similarity scores for each similarity search. The similarity and classifier scores, denoted by S and C respectively, are displayed in the top left corner for each image. The red outlines indicate galaxies which would be visually classified as hosting tidal features, regardless of whether this galaxy is the central object in the image e.g. first row, second column.}
  \label{fig:HSC_sim_search}
\end{figure*}

\subsubsection{UMAP}
\label{sec:UMAP}
Another way to visualise how the model organises the galaxy images in representation space, without having to select a query image, is by using Uniform Manifold Approximation and Projection (UMAP; \citealt{McInnes2018UMAP}). UMAP can be trained directly on image data to produce meaningful clusters. However, here we use it merely for visualisation purposes. UMAP takes our $\sim~$44,000 encoded 128-dimensional representations from the unlabelled dataset as input and reduces them to an easier to visualise 2 dimensional projection. Figure \ref{fig:UMAP_preview} illustrates this 2D projection, created by binning the space into $100~\times~100$ cells and randomly selecting a sample from that cell to plot in the corresponding cell location. To obtain an idea of what attributes galaxy groupings are based on, we hand-select three areas and show zoomed-in versions of these areas around the edges of the figure. Visually, galaxies appear to be grouped both according to their colour, and their size in the cutout. The fact that we can achieve a high performance of tidal feature classification (shown in Section \ref{sec:res_comp}) by simply using a linear classifier on the representations means that the representations created by the self-supervised encoder are meaningful. We also determine whether the scores given to galaxies by the linear classifier are related to the galaxies' positions in the UMAP projection. This is done by binning the space into $40~\times~40$ cells and colouring the cells according to the median classifier score in each bin, shown in the left panel of Figure \ref{fig:UMAP_scores}. We find that the majority of galaxies which were assigned a high classifier score, indicating a high likelihood of tidal features, are located on the left side of the UMAP projection plot. The right panel of Figure \ref{fig:UMAP_scores} shows the median absolute deviation of the classifier score in each bin. This deviation remains low across the plot, only rising slightly at the boundary between high and low classifier scores, indicating that the classifier scores are stable across the UMAP space. The distribution of classifier scores across the UMAP plot reinforces the idea that the encoded representations contain meaningful information about tidal features, but also brings to light a potential bias of our model. The left side of the UMAP projection plot appears to contain the galaxies which cover more of the cutout, indicating a potential tendency towards classifying brighter galaxies that appear larger in the cutouts as having tidal features. This pattern can also be observed in Figure \ref{fig:HSC_sim_search} where galaxies that are bright and cover a large portion of the cutout are assigned high classifier scores despite not exhibiting tidal features. This could be the result of a selection effect introduced by the training set as tidal features are more likely to be visible and obvious for brighter and larger-appearing galaxies and hence these galaxies are more likely to be visually classified as having tidal features. 

\begin{figure*}
    \centering
	\includegraphics[width=0.99\textwidth]{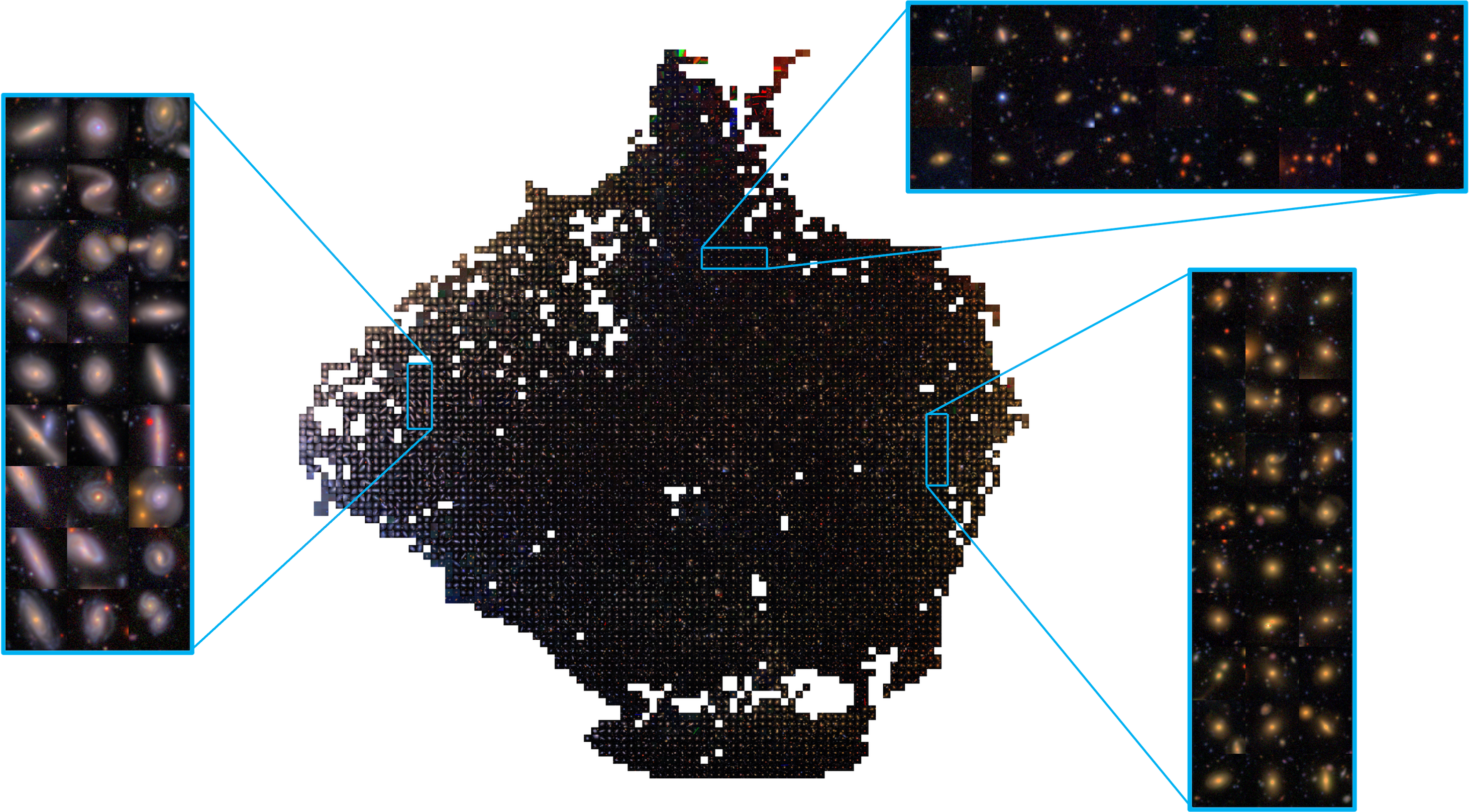}
    \caption{2D UMAP projection of the self-supervised representations. Made by binning the space into $100~\times~100$ cells and randomly selecting a sample from that cell to plot in the corresponding cell location. }
    \label{fig:UMAP_preview}
\end{figure*}

\begin{figure}
    \centering
	\includegraphics[width=0.99\columnwidth]{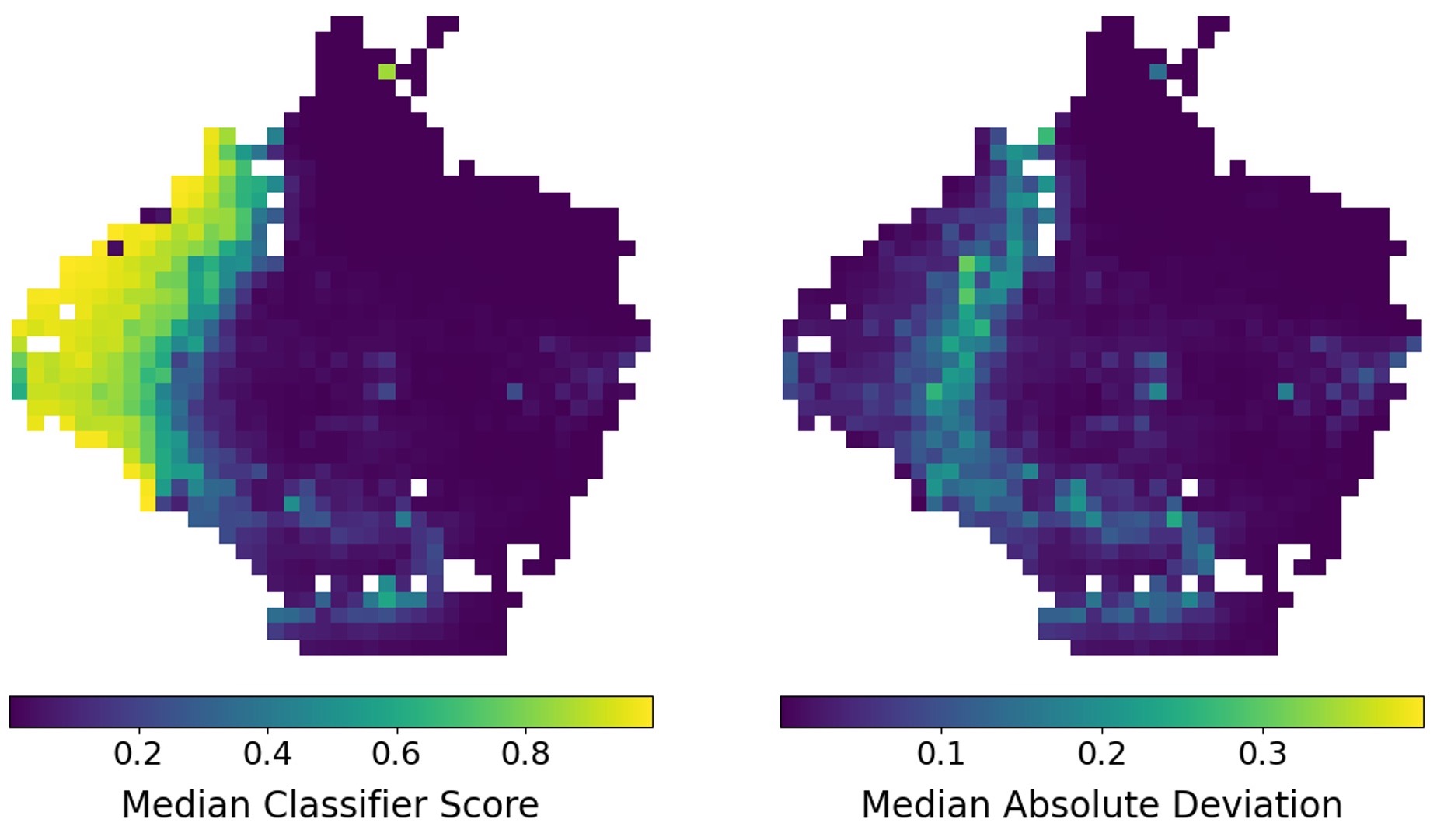}
    \caption{Left: The same 2D UMAP projection as Figure \ref{fig:UMAP_preview}, binning the space into $40~\times~40$ cells, coloured according to the median classifier score (assigned to each galaxy by the linear classifier) in each bin. Bins with less than 5 galaxies are coloured white. Right: The same 2D UMAP projection, this time coloured according to the median absolute deviation of the classifier scores in each bin.}
    \label{fig:UMAP_scores}
\end{figure}

\subsubsection{Dependence of Model Performance on Galaxy Properties}
\label{sec:App}
In this section we investigate whether a link exists between the classifier scores assigned by the model and the distribution of various galaxy properties. The panels labelled A in Figure \ref{fig:dataset_dist} show the distribution of colour, brightness, photometric redshift, and $i-$band radius in our unlabelled dataset ($\sim$44,000 galaxies) and our labelled datasets with tidal features ($\sim$400 galaxies) and without tidal features ($\sim$400 galaxies). We find that galaxies in the labelled tidal feature dataset tend to be brighter and have lower redshifts as well as slightly larger radii than unlabelled galaxies. This can be attributed to the differences in selection between these two samples, described in Section \ref{sec:HSC}. The galaxies in the visually-classified dataset (taken from \citealt{Desmons2023GAMA}) are drawn from the GAMA survey which is apparent-magnitude limited ($r~\leq~19.8$ mag) and has median redshift $z~\sim~0.2$ so galaxies from this dataset tend to be brighter and have lower redshifts as well as slightly larger apparent radii than the broader selection of unlabelled galaxies from the HSC-SSP Ultradeep regions. These selection effects are unavoidable differences between the two samples and the classifier could potentially embed these effects, causing high classifier scores to be assigned disproportionately to brighter and larger-appearing galaxies. 

To investigate whether this is the case we apply the classifier to our large unlabelled dataset to obtain a large sample of galaxies with assigned classifier scores. The panels labelled B in Figure \ref{fig:dataset_dist} show the distribution of the same galaxy properties, this time for galaxies assigned classifier scores $\geq$ 0.8 and galaxies assigned scores $\leq$ 0.2. These panels show that galaxies given high classifier scores share a similar distribution of properties to the galaxies in the labelled tidal feature dataset. Galaxies with high classifier scores tend to be brighter, have lower redshifts, and are more concentrated towards larger apparent radii. This is likely not entirely due to bias within the model, since tidal features are more likely to be visible for brighter, larger, lower redshift galaxies (e.g. \citealt{KadoFong2018HSCTidalFeat,Martin2022TidalFeatMockIm}) However, this result, paired with the galaxies assigned high classifier scores in Figure \ref{fig:HSC_sim_search} despite not possessing tidal feature, is an indication that our model does have some bias towards assigning high classifier scores to galaxies with these properties. We therefore expect that the majority of the contamination in a tidal feature sample created by applying this model would be due to these larger and brighter, low redshift galaxies.

\begin{figure*}
    \vspace{2em}%
    \centering
    \begin{subfigure}[t]{0.99\textwidth}
        \centering
        \includegraphics[width=\textwidth]{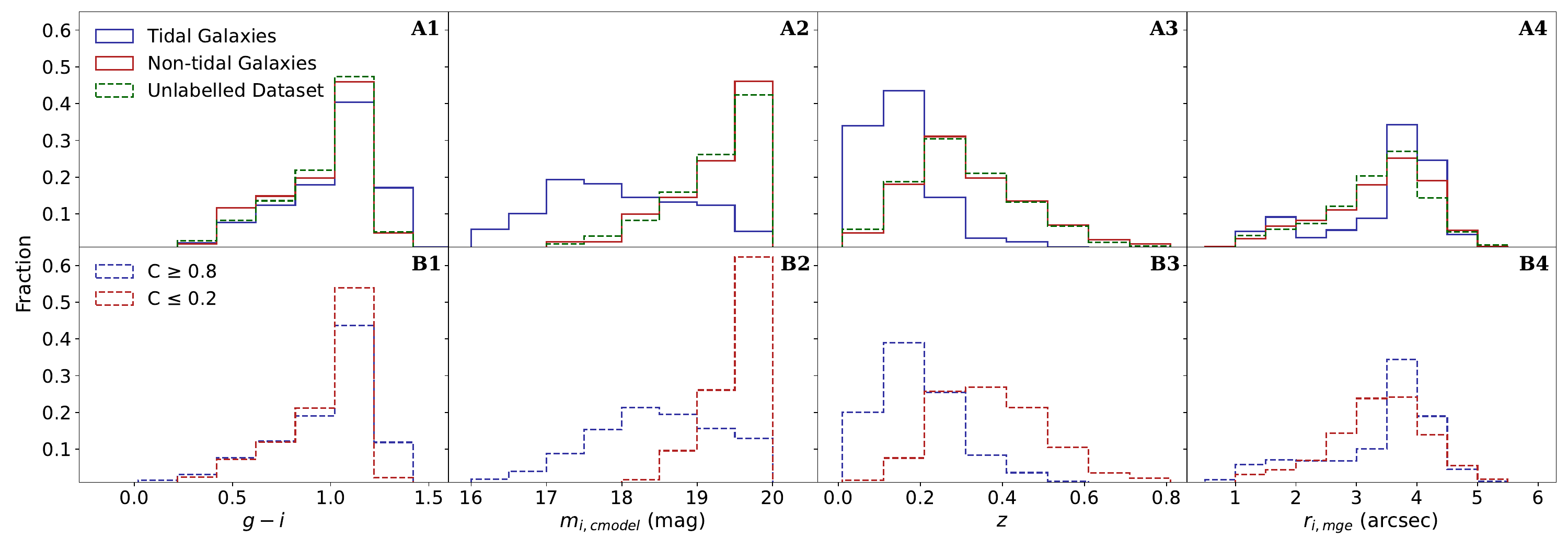}
    \end{subfigure}
    \caption{A Panels: Identical to Figure \ref{fig:Data_prop} -- Distribution of galaxy properties for the labelled datasets of galaxies with known tidal features (blue) and without tidal features (red) and the large unlabelled dataset (dashed green). B Panels: Distribution of galaxy properties for galaxies in the unlabelled dataset assigned a classifier score $\geq$ 0.8 (dashed blue) and $\leq$ 0.2 (dashed red). From left to right: $\it{g-i}$ colour, $\it{i-}$band cmodel fit magnitude, photometric redshift, and $\it{i-}$band MGE fit radius.}
    \label{fig:dataset_dist}
\end{figure*}

To further investigate the effect of galaxy brightness and radius on the output of the model we analyse its performance in five distinct magnitude-radius regions. These regions are chosen empirically to span the full range of magnitudes and radii in the unlabelled dataset, and are shown by the red boxes in Figure \ref{fig:rad_mag_boxes}. Each box covers a region of 0.35 magnitudes in brightness and 0.25 arcseconds in radius. The specific magnitude and radius ranges for each box are defined in Table \ref{tab:mag_box_ranges}, with box 1 containing brighter and larger galaxies, and box 5 containing fainter and smaller galaxies. To conduct our investigation, we randomly choose 200 galaxies from each box, apply the model to obtain a classifier score for each of them, and also perform visual classification (following \citealt{Desmons2023GAMA}) to obtain a true label for each galaxy. The histograms in the top panels of Figure \ref{fig:TF_non_TF_hists} show the distribution of classifier scores in each box, along with the results of our visual classification. Galaxies visually classified as having tidal features are shown in purple, and galaxies without tidal features are shown in blue. We also display the overall tidal feature fraction for each box. From this figure, it is evident that in box 1 the classifier score distribution is heavily skewed towards the right, whereas for the other boxes, the distribution tends to skew left. This confirms our previous findings that high classifier scores are more likely to be assigned to brighter galaxies, with less dependence on radius.

To assess the performance of the classifier in finding tidal features in each of the boxes, we calculate the tidal feature fraction for each classifier score bin by dividing the number of galaxies with tidal features in a bin by the total number of galaxies in that bin. The bottom panels of Figure \ref{fig:TF_non_TF_hists} shows the distribution of this fraction for each of the boxes, as well as the overall tidal feature fraction in the box indicated by the red dashed line in each subplot. The uncertainties on the presented fractional quantities are given for 1$\sigma$, and were estimated using the beta distribution quantile technique to calculate confidence intervals \citep{Cameron2011ConfInt}. We used this method to calculate our confidence intervals as it is well-suited to smaller fractions and performs well for small-to-intermediate sample sizes. In all cases presented in Figure \ref{fig:TF_non_TF_hists}, the tidal feature fraction in the highest classifier score bin is greater than the overall tidal feature fraction for the relevant box. This means that the model is successfully assigning high classifier scores to galaxies with tidal features regardless, of the brightness or radius of the host galaxy. For boxes 2 - 5 the difference between the tidal feature fraction in the high classifier score bins and the overall tidal feature fraction for the box is significant, while for box 1 the result is less significant due to the level of contamination from bright non-tidal galaxies assigned high classifier scores. Although the model's performance is affected by contamination from bright galaxies, the overall success demonstrated for the various ranges of brightness and radius means that the number of galaxies that need to be visually classified to find tidal features in any large dataset will be greatly reduced.

These figures support our statement in Section \ref{sec:mod_eval}, that this model should be used in conjunction with visual classification, to partially automate the detection of tidal features and reduce workload for visual classifiers. Alternatively, if purity is of great importance to the user’s particular science case, another strategy would be to pre-select a set of very uniform galaxies on which to run the model, by applying brightness cuts. Within such a sample the classifier would have less dependency on these galaxy features and would concentrate more on the presence of tidal features.

\begin{figure}
    \centering
    \includegraphics[width=0.95\columnwidth]{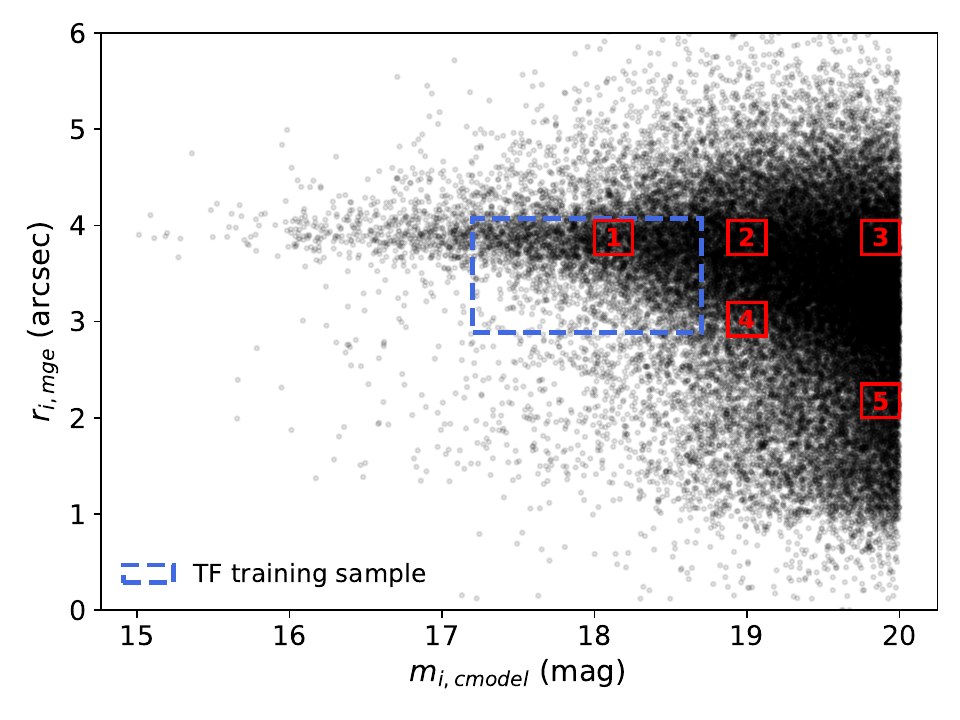}
    \caption{Galaxy $\it{i-}$band MGE radius as a function of $\it{i-}$band cmodel magnitude for the $\sim$44,000 galaxies in the unlabelled dataset. The red numbered boxes show the five regions in which we evaluate the model performance. The blue dashed box indicates the region marked out by the 25$^{\mathrm{th}}$--75$^{\mathrm{th}}$ percentiles of radius and magnitude for the galaxies in the labelled tidal feature dataset.}
    \label{fig:rad_mag_boxes}
\end{figure}

\begin{table}
	\centering
	\caption{$\it{i-}$band cmodel fit magnitude and $\it{i-}$band MGE fit radius ranges for each of the five boxes shown in Figure \ref{fig:rad_mag_boxes}.}
	\label{tab:mag_box_ranges}
	\begin{tabular}{lcc}
		\toprule
		\thead{Box} & \thead{m$_{i,cmodel}$ range (mag)} & \thead{r$_{i,mge}$ range (arcsec)} \\
		\midrule
		1 & 18 -- 18.25 & 3.7 -- 4.05 \\
            2 & 18.875 -- 19.125 & 3.7 -- 4.05 \\
            3 & 19.75 -- 20 & 3.7 -- 4.05 \\
            4 & 18.875 -- 19.125 & 2.85 -- 3.2 \\
            5 & 19.75 -- 20 & 2 -- 2.35 \\
		\bottomrule
	\end{tabular}
\end{table}

\begin{figure*}
    \vspace{2em}%
    \centering
    \begin{subfigure}[t]{0.98\textwidth}
        \centering
        \includegraphics[width=\textwidth]{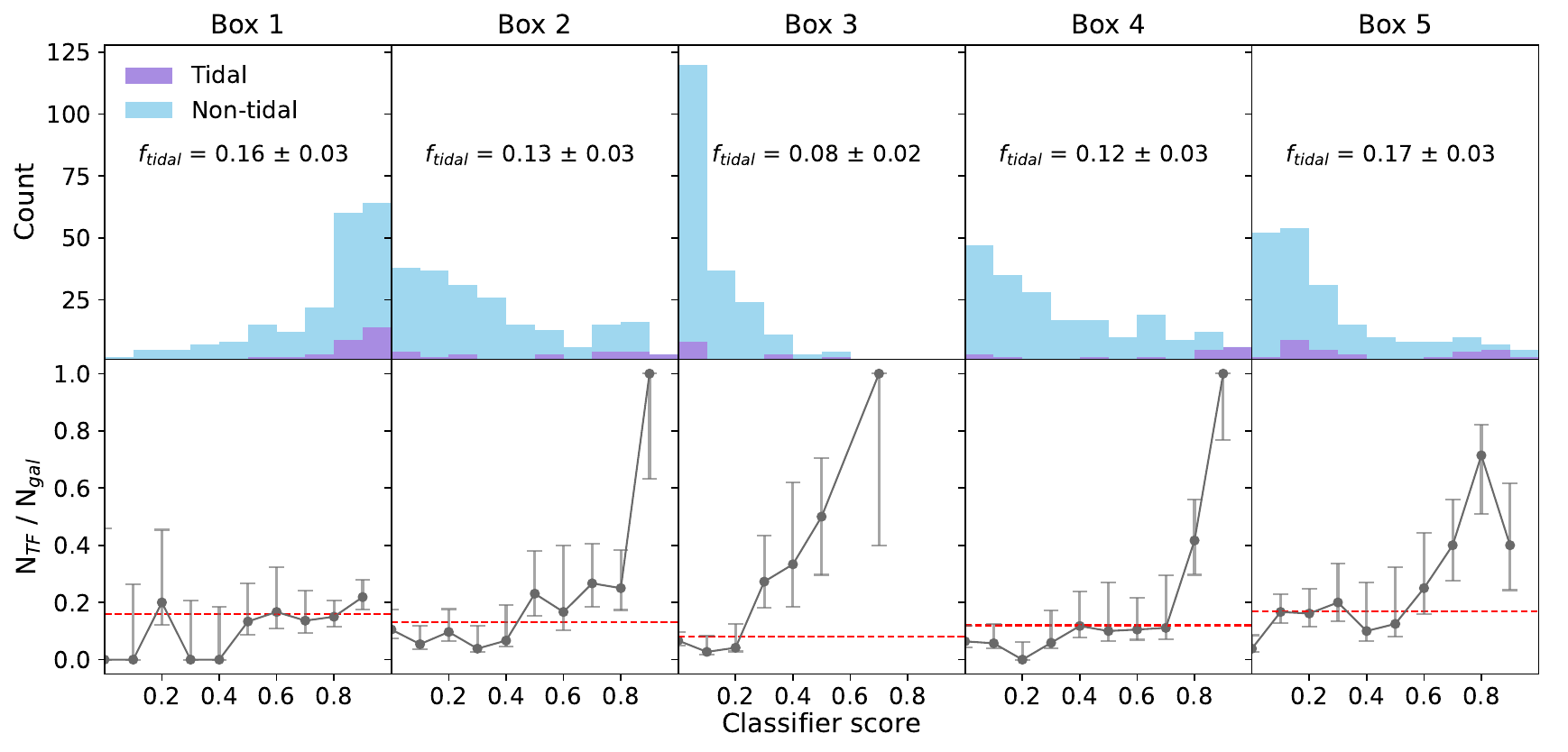}
    \end{subfigure}
    \caption{The upper panels show the distribution of classifier scores assigned by the model to the 200 galaxies in each box. Galaxies visually-classified as having tidal features are shown in purple, and galaxies without tidal features are shown in blue. Each subplot also shows the tidal feature fraction in the respective box, according to visual classification. The lower panels show the Tidal feature fraction in each classifier score bin. Uncertainties are given for 1$\sigma$ and estimated as described in the text. The red dashed line in each subplot indicates the tidal feature fraction for the relevant box, according to visual classification.}
    \label{fig:TF_non_TF_hists}
\end{figure*}

\section{Discussion and Conclusions}
\label{sec:disc}

In this work, we have shown that SSL models composed of a self-supervised encoder and linear classifier can not only be used to detect galaxies with tidal features, but can do so reaching both high completeness (TPR~$=$~0. 94~$\pm$~0.1) for low contamination (FPR~$=$~0.20) and high area under the ROC curve (ROC~AUC~$=$~0.91~$\pm$~0.002). This means that such models can be used to isolate the majority of galaxies with tidal features from a large sample of galaxies, thus drastically reducing the amount of visual classification needed to assemble a large sample of tidal features. One major advantage of this model over other automated classification methods, is that this level of performance can be reached using only 600 labelled training examples, and only drops mildly when using a mere 50 labels for training maintaining ROC~AUC~$=$~0.89~$\pm$~0.01 and TPR~$=$~0.90~$\pm$~0.1 for FPR~$=$~0.2. SSL models are also inexpensive to train, with the encoder needing only $\sim$~30 minutes to train for 25 epochs on a single GPU. The linear classifier, applied to the encoded representations, only required $\sim$~1 minute to train for 50 epochs on a single GPU. This makes SSL models easy to re-train on data from different surveys with minimal visual classification needed. 

Previous works which used SSL models for classification of astronomical objects highlighted a number of advantages of these models compared to fully supervised models. \citet{SteinOct2021SelfSupSim} highlighted the usefulness of being able to use the encoded representations straight from the self-supervised encoder to perform similarity searches based on a single example image in the context of finding strong gravitational lens candidates. \citet{Hayat2021SSMLAstroIms} emphasised the superior performance of their SSL model compared to a supervised model for both galaxy morphology classification, and spectroscopic redshift estimation, particularly when decreasing the number of labels available for training. In this work we find similar advantages of a SSL model compared to a supervised model. When using the models to detect merging galaxies from an SDSS dataset, both models show similar performance when using a larger number of training labels, however, in the regime of fewer training labels, the supervised model performance decreases more drastically than the SSL model. When using the models for the detection of tidal features we find that the SSL model consistently outperforms the supervised model, regardless of the number of labels used for training. Following \citet{SteinOct2021SelfSupSim}, we emphasise the usefulness of being able to perform a similarity search using just the self-supervised encoder and one example of a galaxy with tidal features to find other galaxies with tidal features from a dataset of tens of thousands of galaxies. 

The level of comparison that can be carried out with respect to the results obtained here and other works is limited due to the scarcity of similar works. There are only two studies focusing on the detection of tidal features using machine learning. The first is the work of \citet{Walmsley2019CNNTidalFeat} who used a supervised network to identify galaxies with tidal features from the Wide layer of the Canada-France-Hawaii Telescope Legacy Survey \citep{Gwyn2012CFHTLS}. They used a sample of 305 galaxies with tidal features and 1316 galaxies without tidal features, assembled by \citet{Atkinson2013CFHTLSTidal}, to train a CNN, using augmentations such as image flipping, rotation, and translation to expand their dataset. \citet{Walmsley2019CNNTidalFeat} found that their method outperformed other automated methods of tidal feature detection, reaching 76\% completeness (or TPR) and 22\% contamination (or FPR). Our SSL model, trained on 600 (300 tidal, and 300 non-tidal) galaxies performs considerably better, reaching a completeness of 96\% for the same contamination percentage. \citet{Dominguez2023MockTidalCNN} used this same CNN, designed by \citet{Walmsley2019CNNTidalFeat}, and the dataset presented in \citet{Martin2022TidalFeatMockIm} to also identify galaxies exhibiting tidal features. This dataset consists of $\sim$~6000 synthetic mock HSC-SSP images, including $\sim$~1800 galaxies with tidal features, from the NewHorizon cosmological hydrodynamical simulation at five different surface brightness limits, ranging from $\mu_{r}=$~28~mag~$\rm{arcsec}^{-2}$ to $\mu_{r}=$~35~mag~$\rm{arcsec}^{-2}$. They found that the model was able to successfully identify tidal features, reaching ROC AUC > 0.9 and a completeness of $\sim$~90\% for 22\% contamination. They also separated the performance of the model according to the image surface brightness limits and found that at surface brightness limits close to HSC-SSP UD limits ($\mu_{r}\sim$~30~mag~$\rm{arcsec}^{-2}$) the model reached ROC AUC = 0.93 and completeness of $\sim$~85\% for 22\% contamination. The ROC AUC reached by our model (0.911~$\pm$~0.002) is comparable to that found in \citet{Dominguez2023MockTidalCNN}, however, we reach a significantly higher completeness of 96\% for the same level of contamination. \citet{Dominguez2023MockTidalCNN} also attempted to apply their trained model to real HSC-SSP images, however the model performance decreased significantly, only reaching ROC AUC = 0.64. They noted that this drop in performance could be attributed to the difference in angular resolution between the simulated and real images, or the lack of real background in the simulated images which were used to train their model.

In a similar work, \citet{Bickley2021CNNTidalIllustris} used a CNN to identify recently merged galaxies in a sample of galaxies from the cosmological magnetohydrodynamical simulation IllustrisTNG \citep{Nelson2018IllustrisTNG}. Their dataset was constructed by combining IllustrisTNG images with Canada France Imaging Survey (CFIS; \citealt{Ibata2017CFIS}) image data and metadata to create synthetic CFIS images and the sample consisted of $\sim$75,000 galaxies, including $\sim$37,000 recently merged galaxies. Their CNN achieved good performance, reaching a higher ROC AUC of 0.95 than our SSL model, and a similar completeness of $\sim$95\% for 22\% contamination, although their model required much larger labelled training set compared to our dataset of 600 galaxies. \citet{Bickley2021CNNTidalIllustris} also compared the performance of their CNN to that of visual identification performed by 9 classifiers on a subsample of 200 galaxies. They found that although the CNN recovered a higher fraction of the recently merged galaxies, the visual classifiers were capable of achieving higher purity in their classification. \citet{Bickley2021CNNTidalIllustris} suggested that the best approach to merger identification is to use a CNN to assemble an initial sample of candidates, followed by visual classification to improve the quality of this merger sample. This conclusion is consistent with our proposal for the intended use of our SSL model.

We can also compare one part of our work with that of \citet{Pearson2019DeepLearnMergers} who used a supervised model to identify merging galaxies from a dataset of $\sim$6000 SDSS galaxies. We use the same SDSS dataset to train both a supervised model (based on that of \citealt{Pearson2019DeepLearnMergers}) and our SSL model to compare the performance of the two models. In their work, \citet{Pearson2019DeepLearnMergers} found that their CNN reached a ROC AUC of 0.966 when used on the testing dataset. When trained on the same number of labels, both our supervised and SSL models reached a similar ROC AUC of 0.96 on the testing set. However, when training using only 2\% of the available training data, or 120 galaxies, the supervised model ROC AUC dropped to 0.77, while the SSL model ROC AUC only dropped to 0.85. This shows the advantage of using SSL models, particularly when available training sets are limited in size or have not been assembled yet.

\citet{Bottrell2019DeepLearningMerger} and \citet{Snyder2019IllustrisAutoMergerClass} focused on the classification of galaxy merger signatures using supervised machine learning. \citet{Bottrell2019DeepLearningMerger} used a CNN to classify galaxies according to merger stage, using a series of images from a hydrodynamical simulation. Their model reached high (87\%) accuracy even when using simulated images inserted into real SDSS survey fields to train and test the model. \citet{Snyder2019IllustrisAutoMergerClass} used a random forest algorithm to isolate galaxies that merged or would merge within 250 Myr, using images from the Illustris cosmological simulation \citep{Genel2014Illustris, Vogelsberger2014aIllustris, Vogelsberger2014bIllustris}. Their model reached $\sim$70\% completeness (TPR) for $\sim$30\% contamination (FPR) which is a significantly lower TPR than that reach by our model. However, both the works of \citet{Bottrell2019DeepLearningMerger} and \citet{Snyder2019IllustrisAutoMergerClass} do not focus explicitly on the detection of tidal features and therefore are not directly comparable to the results of our analysis.

In Section \ref{sec:App} we investigated the effect of the differences in galaxy property distributions between our labelled tidal feature and non-tidal feature datasets on the performance of the classifier.  We found one limitation of our model; due to galaxies in the labelled tidal feature dataset being brighter than those in the unlabelled or labelled non-tidal feature datasets, the model was more likely to assign high classifier scores to brighter galaxies. This tells us that in order to ensure model performance remains consistent over the entire range of properties in a galaxy sample, the datasets used to train a model should ideally trace the full range of galaxy properties in the sample to which the model will be applied even for self-supervised models (e.g. \citealt{Walmsley2024Scaling}). However, this model is designed with LSST in mind. While these new data will need a new training sample, ideally spanning the full range of galaxy properties in the LSST dataset, we have shown that fewer labels can be used than for a classical supervised approach. This means this model can be rapidly retrained to be applied to LSST data to assemble a large set of galaxies with tidal features.

In this work we have shown that self-supervised machine learning models can be used to detect galaxies with tidal features from large datasets of galaxies. They can do so reaching both high area under the ROC curve and high completeness (TPR) for low contamination (FPR). To reach good performance, these models do not require large labelled datasets, and can be fully trained using as few as 50 labelled examples. This makes them easy to re-train on new data and therefore, simple to apply to data from different surveys with minimal visual classification needed. Such models can also be used to conduct similarity searches, finding galaxies with similar features given only one labelled example of a galaxy. This can help us understand what image features the model considers important when making links between images, and can be applied to any astronomical dataset to find rare objects. All of these attributes make this SSL model a valuable tool in sorting through the massive amounts of data output by imaging surveys such as LSST, to assemble large datasets of merging galaxies.
The code used to create, train, validate, and test the SSL model, along with instructions on loading and using the pre-trained model as well as training the model using different data can be downloaded from GitHub\footnote{\url{https://github.com/LSSTISSC/Tidalsaurus}}.

\section*{Acknowledgements}
\label{sec:acknowledge}
We thank the anonymous referee for their helpful comments that have improved the quality of this paper.

We acknowledge funding support from LSST Corporation Enabling Science grant LSSTC 2021-5. SB acknowledges funding support from the Australian Research Council through a Discovery Project DP190101943.

The Hyper Suprime-Cam (HSC) collaboration includes the astronomical communities of Japan and Taiwan, and Princeton University. The HSC instrumentation and software were developed by the National Astronomical Observatory of Japan (NAOJ), the Kavli Institute for the Physics and Mathematics of the Universe (Kavli IPMU), the University of Tokyo, the High Energy Accelerator Research Organization (KEK), the Academia Sinica Institute for Astronomy and Astrophysics in Taiwan (ASIAA), and Princeton University. Funding was contributed by the FIRST program from the Japanese Cabinet Office, the Ministry of Education, Culture, Sports, Science and Technology (MEXT), the Japan Society for the Promotion of Science (JSPS), Japan Science and Technology Agency (JST), the Toray Science Foundation, NAOJ, Kavli IPMU, KEK, ASIAA, and Princeton University. 
This paper makes use of software developed for Vera C. Rubin Observatory. We thank the Rubin Observatory for making their code available as free software at http://pipelines.lsst.io/.
This paper is based on data collected at the Subaru Telescope and retrieved from the HSC data archive system, which is operated by the Subaru Telescope and Astronomy Data Center (ADC) at NAOJ. Data analysis was in part carried out with the cooperation of Center for Computational Astrophysics (CfCA), NAOJ. We are honored and grateful for the opportunity of observing the Universe from Maunakea, which has the cultural, historical and natural significance in Hawaii.\\ 
\section*{Data Availability}
\label{sec:data_avail}
 
 All data used in this work is publicly available at \url{https://hsc-release.mtk.nao.ac.jp/doc/} (for HSC-SSP data).


\bibliographystyle{mnras}
\bibliography{bibs} 





\bsp	
\label{lastpage}
\end{document}